# Ultrasound and Magnetic Resonance Elastography of the Neck: Biomechanical Basis, Clinical Applications and Methodological Challenges

Rashedul Hasan,[a] Anna Yang,[b] Kishani Kandasamy,[c] Hossein Bahreinizad,[b] Md Ashikuzzaman,[c] Suman K. Chowdhury,[b] Shahed Mohammed[a,*]

[a]University of North Carolina at Charlotte, Electrical and Computer Engineering, Charlotte, NC

[b]University of Florida, Industrial and System Engineering, Gainesville, FL

[c]University of Missouri-Kansas City, Electrical and Computer Engineering, Kansas City, MO

*Corresponding Author: smoham30@charlotte.edu

## Abstract

Many neck disorders, from chronic neck pain and headache to thyroid and nodal malignancy and radiation-induced fibrosis, both arise from and alter the mechanical properties of neck tissues, yet clinical assessment still relies on morphology, palpation and symptoms. Ultrasound and magnetic resonance elastography (MRE) now allow these properties to be measured in vivo, but the neck literature differs widely in methodology and remains dispersed across modalities and tissues. This review brings together the biomechanical basis of neck tissue behavior, the clinical conditions that alter it, and the principles and applications of ultrasound elastography and MRE in the neck. We show that ultrasound elastography has demonstrated diagnostic value for thyroid nodules and cervical lymph nodes, while neck MRE has been shown to be feasible but remains at an early stage. Across both modalities, reported values depend strongly on fiber orientation, posture, muscle activation and acquisition conditions. We therefore propose reporting recommendations and discuss how measurements beyond stiffness, including anisotropy, viscosity and nonlinearity, could improve disease management and inform computational models of the neck.

## 1. Introduction

The neck is the sole, mechanically active link between the head and the trunk, where the neck musculoskeletal system provides mobility and load support while protecting the airway, digestive tract and neurovascular structures [4, 5]. Emerging evidence indicates that intrinsic tissue mechanical properties can both contribute to and reflect many neck disorders. Firstly, sustained occupational loading and prolonged non-neutral postures increase mechanical demand on the cervical musculature and are associated with muscle stiffening and chronic neck pain [6-8]. Additionally, fibrosis and malignancy stiffen thyroid nodules and lymph nodes and radiotherapy leaves fibrotic tissue that impairs swallowing [9-11]. Yet diagnosis still relies on morphological imaging, palpation and self-reported symptoms, approaches unable to capture the mechanical dysfunction that likely contributes to pain, functional impairment and disease progression [12]. Quantitative, in vivo measurement of neck tissue mechanical parameters could enable objective diagnosis and advance understanding of the pathogenesis of many neck diseases.[13].

Over the past three decades, elastography has made it possible to measure tissue mechanical properties in vivo. Ultrasound strain elastography estimates relative stiffness from tissue deformation under quasi-static compression, whereas shear wave elastography derives stiffness from the speed of shear waves generated by acoustic radiation force [14, 15]. Magnetic resonance elastography (MRE) instead image externally induced shear waves throughout a tissue volume [16]. However, the mechanical complexity of neck tissues poses practical and methodological challenges for both approaches. Neck muscles are anisotropic, so measured stiffness depends on wave propagation direction relative to the fibers [17, 18]. Measured stiffness also changes substantially with posture and muscle activation, both of which must be controlled during measurement [19, 20]. Many neck targets are small, superficial and close to bone, which complicates wave delivery and transducer contact [3, 18]. Reported values further depend on acquisition conditions such as probe pressure, excitation frequency, inversion algorithm and imaging system [21-23]. Without careful control and reporting of these factors, the mechanical parameters obtained are difficult to interpret or compare across studies. At the same time, characterizing tissue mechanics beyond apparent stiffness, by also measuring anisotropy, nonlinearity and viscosity, could provide a more complete description of tissue behavior [24-26]. This could be impactful for both disease management and the development of biomechanical simulations and digital twins of the neck [27].

Previous reviews have addressed MRE of muscle or the lumbar spine [13, 28, 29], or ultrasound of neck muscle in pain [30, 31]. None has examined neck elastography across both modalities, linked neck pathologies to measurable mechanical parameters, or considered how in vivo data could inform cervical models [32, 33]. This narrative review addresses these gaps by bringing together biomechanics, clinical evidence and imaging methodology, an approach intended to serve the clinicians, imaging scientists and biomechanical engineers who together drive this field. The review is organized as follows. Section 2 describes the literature search. Section 3 outlines the biomechanical behavior of neck tissues and its biological basis, and Section 4 reviews the clinical conditions that alter these properties. Section 5 describes how neck biomechanics are currently assessed, through clinical examination and computational modeling, and the limitations of these approaches. Sections 6 and 7 cover the principles and neck applications of MRE and ultrasound elastography, respectively. Section 8 compares the two modalities and considers measurements beyond stiffness and their implications for computational modeling. Section 9 discusses technical and methodological challenges and proposes reporting recommendations, and Section 10 concludes.

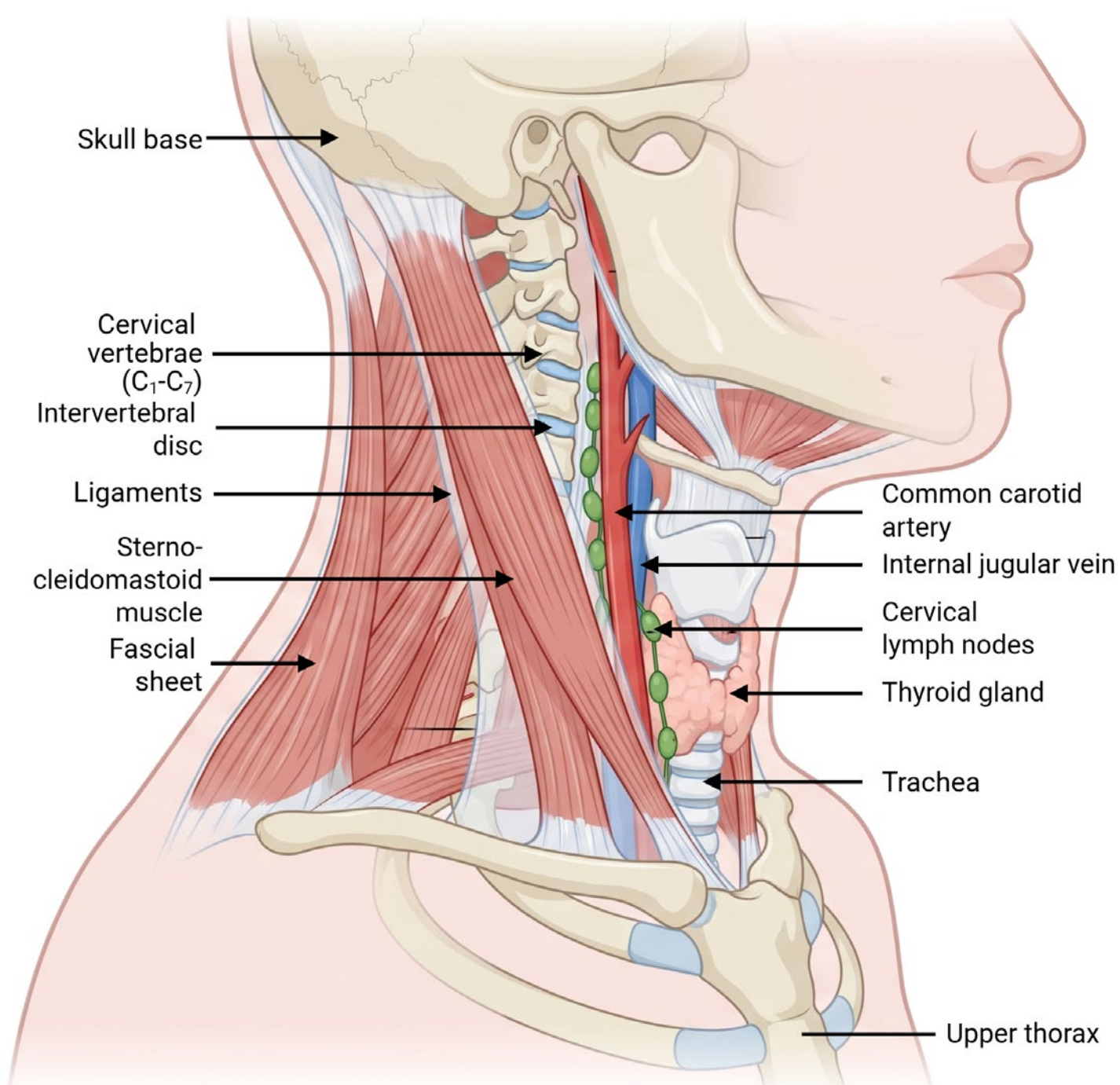

**Figure 1: Mechanical and geometric diversity of neck tissue**

## 2. Information sources and search strategy

A search of articles was performed using PubMed, supplemented by a hand search using Google Scholar and the bibliographies of the included articles. The date of the last search was 6 October, 2026. For the clinical significance section, each disease name was combined with elastography (e.g., chronic neck pain

AND elastography). For the MR elastography section, method and biomarker names were combined with magnetic resonance elastography (e.g., pneumatic driver AND magnetic resonance elastography; viscosity AND elastography), and current neck MRE studies were identified using neck MR elastography; studies using shear wave or strain elastography were excluded from this section. For the ultrasound elastography sections, [search terms] were used, and MRE studies were excluded. Across all sections, application studies of non-cervical head structures, such as the brain and ear, were excluded, whereas studies of other organs were retained where they described elastography methods or biomarkers. No filters were applied for publication date or article type. Background material on neck biomechanics, clinical conditions and elastography principles (Sections 3–5) was drawn from key literature selected by the authors, supplemented by studies of other organs where neck-specific evidence was lacking.

## 3. Musculoskeletal Biomechanics of the Neck

The neck contains many mechanically and geometrically distinct tissues (Fig. 1): vertebrae, intervertebral discs, facet joints, ligaments, muscles and fascia, as well as glands, lymph nodes and vessels [5]. The mechanical behavior of the neck as a whole arises from how these tissues interact with geometry, loading and muscle activation [34-36]. This section traces how these properties arise, working from the whole system down to individual tissues. Section 3.1 describes the cervical spine as a coupled mechanical system, and Section 3.2 describes how head loads are transmitted through its tissues. Section 3.3 distinguishes whole-neck structural stiffness from tissue material properties. Sections 3.4 and 3.5 then examine the passive and active sources of mechanical resistance. Finally, section 3.6 explains how viscoelasticity, anisotropy and heterogeneity follow tissue microstructure.

### 3.1 The Neck as a Coupled Mechanical System

The cervical spine extends from the craniocervical junction to the cervicothoracic junction. It is commonly described as a series of functional spinal units, each made up of two adjacent vertebrae, the intervertebral disc between them, the facet joints and the supporting ligaments (Fig. 2). Because these units are linked in series, the mechanical behavior of the cervical spine comes from their combined response rather than from any single level [37]. Flexion, extension, axial rotation, and lateral bending (Fig. 3) are therefore distributed across the cervical spine and often occur as coupled motions [35, 38].

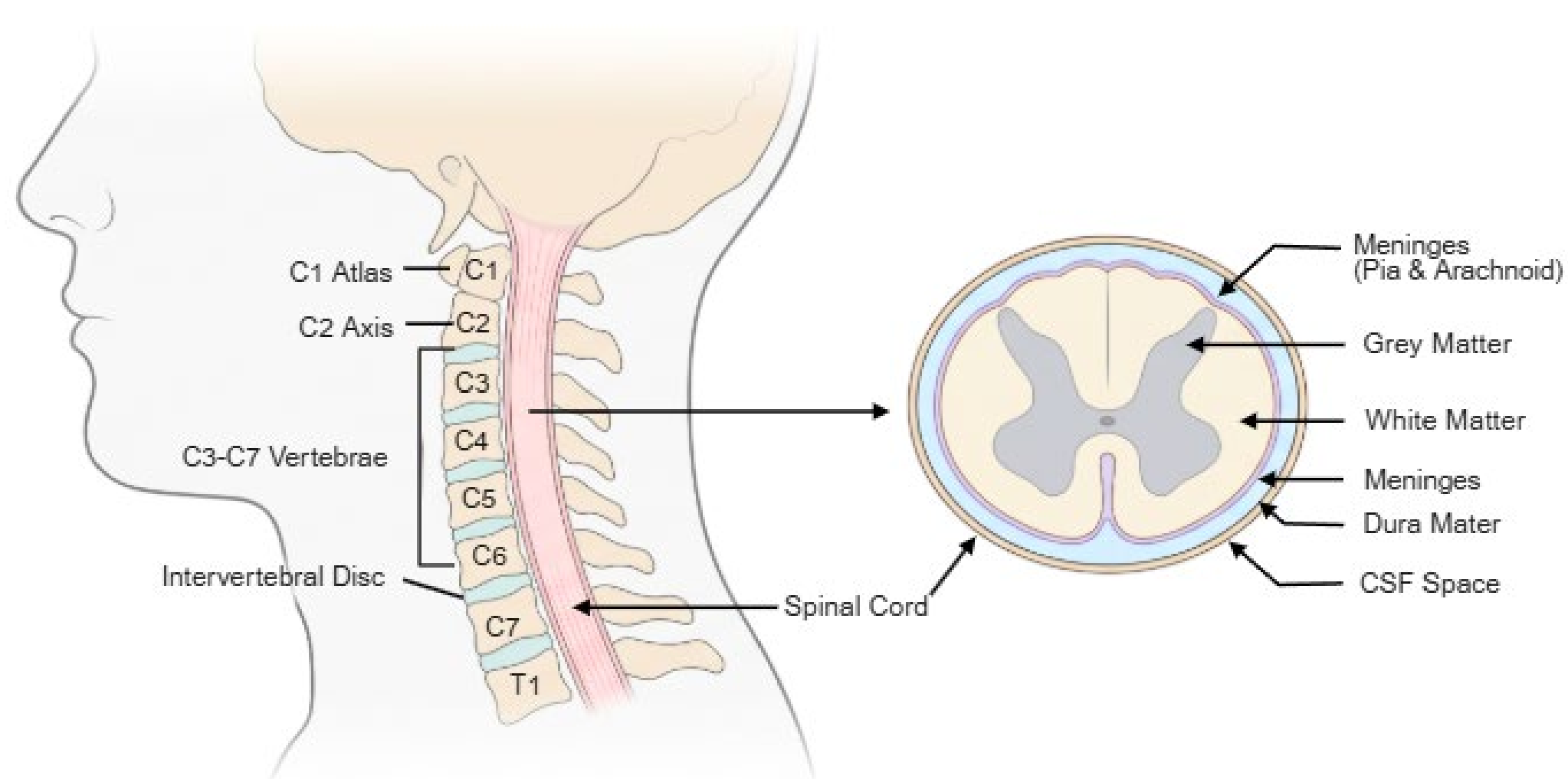

**Figure 2: Cervical spine anatomy, showing the vertebrae (C1–T1), intervertebral discs and spinal cord, with an axial cross-section of the cord.**

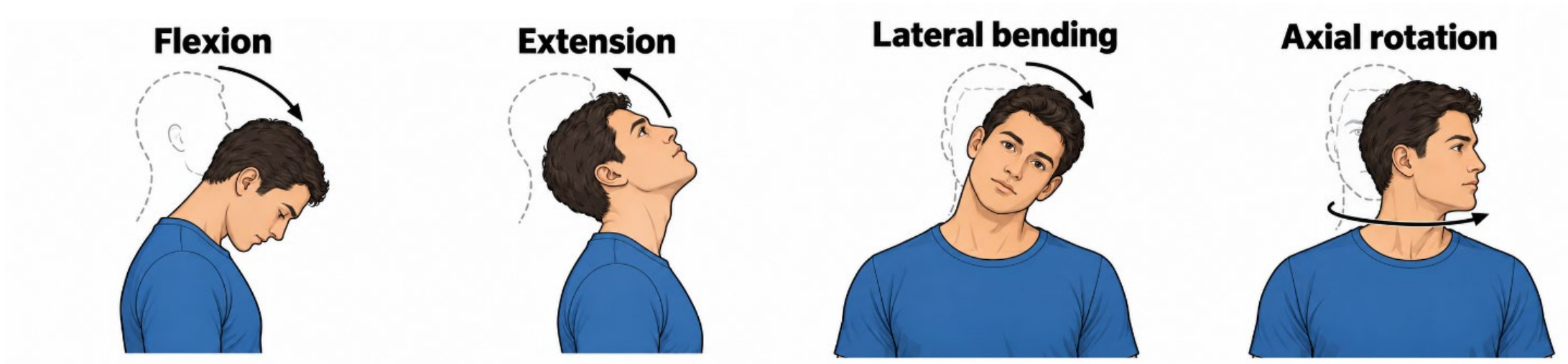

**Figure 3: Primary Movements of the Cervical Spine: Flexion, Extension, Lateral Bending, and Axial Rotation**

Motion is distributed differently in the upper and lower cervical spine. The upper cervical spine, from C0 to C2, is specialized for large movements of the head. The atlanto-occipital joint (C0-C1) contributes primarily to flexion and extension. The atlanto-axial joint (C1-C2) provides for more than half of total cervical axial rotation [39, 40]. Below C2, in the subaxial spine, motion is distributed across several levels depending on the movement direction. Within this region, flexion–extension is greatest at the mid-to-lower segments (C4–C6). Lateral bending is also shared across the subaxial levels and is typically accompanied by axial rotation in the same direction, while these levels contribute only small amounts of axial rotation on their own [40-43]. Overall, the upper cervical spine orients the head, while the subaxial spine distributes motion and controls the position of the head relative to the trunk.

The mobility of cervical system is balanced by passive restraints that limit excessive translation and rotation, protecting neural structures and maintaining stability. The facet joints guide segmental motion and resist shear, particularly anterior shear in the subaxial spine [44, 45]. The facet capsules and spinal ligaments become taut as deformation progresses beyond the relatively lax neutral region [45, 46]. Therefore, the mechanical behavior of the reflects a balance between mobility and restraint. It depends on posture, joint position and muscle activation as well as on the properties of the individual tissues.

### 3.2 Load Transmission and Head Support

Beyond permitting motion, the cervical spine must support the head and transmit its loads to the thorax. These loads come not only from the weight of the head but also from the muscle forces needed to hold the head in posture and balance [47]. The gravitational moment of the head depends on the horizontal distance between the center of mass and the cervical center of rotation. As the head moves away from a neutral posture, this moment and the corresponding muscle forces change, altering the loads transmitted through the cervical spine [6, 48]. Posture also affects how these loads are distributed. For example, compared with a normal lordotic curvature, a straightened cervical spine alters stress concentrations within the intervertebral discs and facet joints [49].

These loads are transmitted through both the anterior and posterior elements of the cervical column. Anteriorly, compressive load is transmitted through the vertebral bodies and intervertebral discs, which form one of the principal load-bearing pathways of the cervical spine [50]. The intervertebral discs then carry and redistribute compressive forces between adjacent vertebral bodies while still permitting relative motion between cervical segments [51]. Posteriorly, the facet joints provide an additional load-sharing pathway that help guide and constrain intervertebral motion [44, 45]. Each intervertebral disc consists of an outer annulus fibrosus (AF), a central nucleus pulposus (NP), and cartilaginous endplates that connect the disc to the adjacent vertebral bodies [5].

The tissues along these load pathways derive their mechanical behavior from their composition and structure. In the vertebral bodies, both the outer cortical shell and the inner porous trabecular network are made of mineralized collagen matrix [5]. The mineral phase provides rigidity, while the collagen provides tensile resistance and toughness. Because this matrix is structurally organized, the mechanical properties of bone differ with loading direction [52]. The intervertebral discs combine different constituents for a different role. Proteoglycans in the nucleus pulposus retain water and generate a swelling pressure that supports compressive load, while collagen in the annulus fibrosus resists tension and restrains disc expansion [53]. Under compression, the solid matrix deforms and fluid pressure rises. Fluid then gradually redistributes, allowing further deformation over time, so the disc's response depends on loading duration and on how freely fluid can move. The articular cartilage of the facet joints follows a similar principle. It consists of a collagen network, proteoglycans and interstitial water, and fluid pressure supports part of the compressive load. As fluid redistributes under sustained deformation, the tissue undergoes stress relaxation [54]. Its collagen fibers also change orientation with depth. They run mainly parallel to the surface in the superficial zone and become more perpendicular to the underlying bone in the deep zone, giving the cartilage depth- and direction-dependent mechanical properties [55].

In addition to these load-bearing structures, spinal ligaments and facet joint capsules elongate and develop tensile forces as adjacent vertebrae translate or rotate, restricting further motion [56, 57]. Such resistance depends on fiber composition and organization. In collagen-rich ligaments, stretching straightens the crimped bundles and progressively recruits fibers, producing nonlinear stiffening [58]. In facet capsules, collagen orientation varies across the tissue, differentiating regional responses of deformation [59]. In this way, load-bearing structures and passive restraints work together to transmit head loads to thorax while keeping intervertebral motion within its feasible range.

### *3.3 Passive Stiffness and Active Muscle Tone*

In biomechanics, neck stiffness describes the resistance of the cervical system to displacement or rotation under an applied force or moment. Translational stiffness is determined from the force-displacement relationship and expressed in N/mm, whereas rotational stiffness is determined from the moment-rotation relationship and expressed in N·m/degree or N·m/rad, depending on the angular unit used [60]. This structural stiffness reflects the combined effects of tissue material properties, anatomical geometry, joint configuration, and muscle activation rather than the mechanical property of an individual tissue [61, 62].

The mechanical response of the neck in vivo reflects the combined effects of passive tissue stiffness and active muscle force. Passive stiffness arises from structures such as the intervertebral discs, ligaments, facet joint capsules, fascia, and the passive components of muscle, whereas active resistance is generated through cervical muscle contraction and neuromuscular control [62]. During movement or posture, both active muscle tone and passive stiffness act simultaneously, contributing to the mechanical response of the neck differently. Even when the head is stationary, cervical muscles may remain active to maintain posture and balance the gravitational moment of the head. During voluntary movement or external loading, changes in muscle recruitment can therefore alter effective neck stiffness without changing the underlying tissue structure [19].

Whole-neck structural stiffness differs from the regional shear modulus estimated by MRE, which typically characterizes tissue resistance to shear deformation. Both passive tension and active muscle contraction can influence this measurement. A higher measured shear modulus does not necessarily indicate a change in the tissue structure of muscles; it may also reflect increased muscle activation or passive tension. Multifrequency MRE studies of skeletal muscle have shown that contraction increases measured shear elasticity compared with the relaxed condition [24]. Therefore, head position, muscle activation, and passive stretch should be standardized during scanning, and MRE-derived tissue stiffness should not be treated as a direct measure of whole-neck structural stiffness.

### *3.4 Passive Mechanical Contributions*

The passive tissues introduced in Section 3.3 do not resist deformation with a fixed stiffness. Their resistance changes with hydration, and with the amount, rate and duration of loading [45, 51, 57]. Because the disc carries part of its load through fluid (Section 3.2), its stiffness depends on loading rate and history. Faster loading increases the apparent compressive modulus because fluid has less time to leave the tissue. Sustained loading, by contrast, causes gradual fluid loss and a progressive change in stiffness [63].

The resistance of the cervical ligaments and facet capsules depends on both how far and how fast they are stretched. Their tension increases with elongation, which limits motion near the end of the physiological range. In the facet capsule, greater joint distraction produces higher capsular strain and tensile force, and excessive stretch leads to yielding and damage [57]. Higher strain rates increase stiffness and failure force, particularly in the posterior longitudinal ligament, ligamentum flavum and capsular ligament [64]. Similar rate-dependent changes in stiffness, failure load and energy absorption have been reported for the anterior longitudinal ligament and ligamentum flavum [65].

Muscle also resists stretch when it is not contracting. Within the fibers, titin resists sarcomere elongation. Outside the fibers, the collagen-containing endomysium, perimysium and epimysium surround individual fibers, fascicles and the whole muscle; this extracellular matrix provides passive resistance and transmits force [66, 67]. As the muscle is stretched, its collagen bundles reorient and increasingly bear load, producing nonlinear resistance [68]. The organization of these bundles also makes this resistance direction-dependent, and fascia adds further connective-tissue resistance (Standring, Ellis et al. 2005).

### *3.5 Active Mechanical Contributions of Cervical Muscles*

Cervical muscles add an active mechanical component to the neck by generating forces for head support, posture, and movement. This force arises within the sarcomere, where myosin cross-bridges attach to actin filaments. Attached cross-bridges both generate force and resist changes in filament position [69], so muscle stiffness rises as activation increases and more cross-bridges are engaged. In the neck, the shear modulus of the extensor muscles increased from rest to isometric head lifting [19] measured five neck extensor muscles at rest and during isometric head lifting and found that their shear modulus increased with active contraction. The shear modulus of the sternocleidomastoid likewise increased with activation during isometric neck torque production [70]. Unlike the passive tissues, whose properties change with loading history and tissue condition, muscle can therefore change its mechanical properties almost immediately through activation.

The active contribution also differs between muscles, because each has a different moment arm and force-generating capacity in flexion–extension, lateral bending and axial rotation [71, 72]. Deep extensors such as the semispinalis cervicis and multifidus lie close to the vertebral column and mainly provide segmental support and control. Larger superficial muscles contribute more to generating head and neck motion [71, 73]. The active mechanical contribution of the cervical musculature therefore depends not only on how strongly individual muscles contract, but also on which muscles are recruited and how their forces are coordinated. Combined with the passive behaviors described in Section 3.4, this means that neck tissues cannot be characterized by a single stiffness value, as Section 3.6 discusses.

### *3.6 Viscoelasticity, Anisotropy, and Heterogeneity*

The behaviors described in Sections 3.2–3.5 mean that cervical tissues cannot be represented by a single, uniform elastic stiffness. Three properties, each rooted in tissue microstructure, account for most of this complexity: viscoelasticity, anisotropy and heterogeneity.

Viscoelasticity is the dependence of a tissue's response on the rate and duration of loading. It arises from fluid flow through proteoglycan-rich matrices and from the time-dependent behavior of collagen and muscle. It underlies the rate effects and stress relaxation already described for the discs, cartilage, ligaments and passive muscle (Sections 3.2 and 3.4). Under cyclic loading, the same behavior makes the mechanical response depend on frequency.

Anisotropy is direction-dependent mechanical behavior, and it arises wherever fibers are aligned. In skeletal muscle, the shear modulus differs parallel and perpendicular to the fibers [26, 74]. The ligaments, facet cartilage and muscle extracellular matrix likewise owe their directional behavior to how their collagen is organized (Sections 3.2 and 3.4). The annulus fibrosus is anisotropic because of its lamellar collagen architecture [75].

Heterogeneity refers to regional variation in mechanical properties within the same tissue or anatomical structure. In the cervical disc, the nucleus pulposus, annulus fibrosus and cartilaginous endplate differ in swelling pressure, permeability, porosity and other biphasic properties [76]. Among the neck muscles, resting stiffness differs between individual muscles and generally increases from superficial to deep layers [19]. Mechanical properties therefore vary with location, even within a single anatomical region of the neck. Because these properties are determined by tissue composition and structure, diseases that alter composition, such as fibrosis, inflammation, edema or malignancy, also alter tissue mechanics, as reviewed in Section 4.

## 4 Clinical Significance of Neck Elastography

Because tissue mechanical properties are determined by composition and structure (Section 3), diseases that alter muscle activation, collagen content, cellularity or fluid balance also alter tissue mechanics. This section reviews the neck conditions in which such changes occur. Sections 4.1–4.3 cover musculoskeletal conditions, in which stiffness changes mainly through altered muscle activation and neuromuscular control: chronic neck pain, headache disorders, and occupational and postural loading. Sections 4.4–4.6 cover conditions that change tissue composition: malignant remodeling of the thyroid and cervical lymph nodes, salivary gland disease, and radiation-induced fibrosis after head and neck cancer treatment. For each condition, the section outlines its prevalence and clinical burden and the mechanisms by which it alters tissue mechanics.

### *4.1 Chronic Neck Pain*

Neck pain is among the most common conditions seen in clinical practice, with an estimated global annual prevalence of approximately 30% [77]. Chronic non-specific neck pain (CNSNP) arises from mechanical stresses such as poor neck posture and from myofascial disorders, often despite normal MRI findings. Myofascial trigger point, which are tender spots within taut bands of muscle, are particularly common: in a cohort of 126 patients, 88.9% showed at least one [78].

Several mechanisms may explain why muscles stiffen in CNSNP. They act through both the passive and active components of muscle described in Sections 3.4 and 3.5. At the motor endplate, tender muscles are thought to release excess acetylcholine, causing sustained contraction, local ischemia and sarcomere shortening. The inflammatory mediators released as a result keep the cycle going, so that irritable muscles become stiff, spastic and shortened [79]. Sustained postural loading may also contribute: repeated or prolonged loading of the cervicothoracic muscles is thought to gradually alter their resting tone [80]. Motor control is impaired as well. Electromyographic studies show that the deep cervical flexors activate late during rapid arm movement, which means the normal feedforward stabilization fails. Meanwhile, the superficial sternocleidomastoid and scalene muscles show increased activity that does not relax normally, which is thought to compensate for the weak deep muscles [81]. Biopsy studies also show a shift from slow-twitch type I fibers toward fast-twitch, glycolytic type II fibers.

Vascular and neural mechanisms have also been proposed. Higher trapezius stiffness is associated with reduced blood flow in the transverse cervical artery, particularly in people who report neck and upper-back stiffness. Reduced perfusion may promote ischemic muscle spasm and a local build-up of free radicals, lactic acid and inflammatory cytokines, while lowering tissue pH and oxygenation [82]. In the neural pathway, chemical mediators released during inflammation, ischemic contraction or fatigue stimulate group III and IV afferents. These afferents excite γ-motor neurons and increase muscle spindle output, raising muscle stiffness [9]. Together, these mechanisms predict increased muscle stiffness in CNSNP, arising from both altered activation and changes in passive muscle properties.

### *4.2 Headache Disorder*

Cervicogenic headache is a secondary headache disorder, meaning it is caused by another condition, in this case dysfunction of the cervical spine and the soft tissues around it. It affects 0.4–2.5% of the general population and accounts for 15–20% of chronic headache presentations. Pain from the neck is referred to the head through the trigeminocervical complex. This region extends from the trigeminal nucleus caudalis to the C2–C3 segments, and there afferents from the upper cervical nerves (C1–C3) converge with those of the trigeminal nerve [83, 84]. The trapezius and sternocleidomastoid also linked to this pathway: fibers of the spinal accessory nerve, which supplies both muscles, connect with the upper cervical nerve roots and the descending tract of the trigeminal nerve. This network is thought to explain the well-recognized patterns of pain referred from these muscles to the face and head [85]. Structurally, the semispinalis capitis, longissimus capitis, longus colli and suboccipital muscles show altered size and thickness, with fat infiltration and atrophy. These changes affect both sources of muscle resistance described in Section 3. Altered activation and spasm change active stiffness, while fat infiltration and atrophy change the composition of the passive tissue and increase its heterogeneity.

### *4.3 Occupational and Postural Loading*

Occupational and postural loading is the sustained or repeated loading of the cervicothoracic muscles during work and daily activities. Forward-head posture, prolonged neck flexion while using devices, sustained sitting and computer work all increase the demand on the cervical extensors, upper trapezius, levator scapulae and scapular stabilizers. As described in Section 3.2, moving the head forward increases its gravitational moment about the cervical spine, and these muscles must generate more force to balance it. When exposure exceeds what the tissue can tolerate and recover from, this demand can lead to fatigue, impaired relaxation and increased stiffness. Repeated or prolonged loading may also gradually change resting muscle tone. However, a posture observed at a single point in time does not reliably predict chronic neck pain. It does not show how long or how often the exposure occurs, whether there is time for recovery, or other physical and psychosocial factors [80].

Such loading is common and causes a substantial clinical burden. Among 359 hospital workers, nearly 75% reported neck and shoulder pain, known in Japan as *katakori*. Computer work of more than 6 h per day was associated with higher odds of symptoms (adjusted OR 2.48) [7]. Sonographers are a particularly exposed group. During clinical scans, their necks were flexed, bent sideways or rotated by more than 20° for 37% of scanning time, and static trapezius activity reached a level rated as medium risk for shoulder–neck disorders [86]. In a Swedish survey of 291 female sonographers, 58% met the criteria for neck or shoulder pain. This pain was associated with high mechanical exposure, high job demands, computer-related eye complaints and headache [87]. In 92 sonographers, resting stiffness of the cervical and axioscapular muscles did not differ between those with no, mild or moderate-to-severe work-related neck disability [88].This contrasts with task-dependent changes in office computer work, where upper trapezius stiffness varied with chair recline, head-and-neck position and sex [8].

Together, these findings suggest that occupational loading may affect how muscles behave during work more than it affects their resting properties.

### *4.4 Malignant Remodeling of Thyroid gland and cervical lymph node*

Malignant remodeling is the structural reorganization of a gland or lymph node as a tumor develops within it. Malignant lesions have greater cellularity, more abundant extracellular matrix, and increased vascularity and interstitial pressure, all of which can alter the mechanical properties of the tissue [89]. These changes act on the constituents described in Section 3. Collagen deposition stiffens the matrix, while raised fluid pressure can stiffen tissue even when the matrix itself is unchanged (Section 6.2.6).

In the thyroid, nodules are palpable in about 5% of people but are detected by ultrasound in up to 50% of the general population, and their stiffness rises with fibrotic content and with the expression of galectin-3 and fibronectin-1, two proteins involved in cell–matrix interaction [11]. This reflects a wider principle whereby cancer and fibrosis change the elastic behavior of tissue, so that the elastic modulus of malignant tissue may be up to an order of magnitude higher than that of benign tissue[90], largely because desmoplastic and fibrotic changes make malignant lesions firmer than benign lesions and adjacent normal tissue [91]. The extent of this stiffening differs between tumor types: collagen-rich fibrosis is frequent and often widespread in papillary thyroid carcinoma but less frequent and less extensive in follicular neoplasms, which may partly account for the inconsistent elastography performance reported across thyroid cancers[92].

In head and neck cancer, cervical lymph node metastasis is a major prognostic factor and the primary reason for treatment failure [93], and nodal metastasis was confirmed in about 81% of patients undergoing therapeutic neck dissection and 69% of those undergoing prophylactic dissection *[94]*. In the oral cavity, where squamous cell carcinoma accounts for about 90% of malignancies, the dense submucosal lymphatic network predisposes these tumors to micrometastatic and bilateral nodal spread [95]. The mechanical properties of an affected node depend on its underlying pathology: lymphomatous nodes typically appear soft, whereas infectious nodes containing scarring, calcification or necrosis, as in tuberculous lymphadenitis, may appear stiffer or heterogeneous *[96]*.

### *4.5 Salivary Gland Disease*

Salivary gland disease can alter tissue mechanics either diffusely, across the whole gland, or focally, as a discrete lesion. The main diffuse example is primary Sjögren's syndrome, a systemic autoimmune disease that targets the salivary glands. Its reported prevalence varies widely with study design and disease definition, from 0.01% to more than 3% of the general population. The first US population-based study estimated it at 2.2–10.3 cases per 10,000 inhabitants, depending on whether classification criteria or physician diagnosis was used [97]. As the disease advances, lymphocytes progressively infiltrate the gland, the ductal epithelium becomes hyperplastic, and fibrosis develops [98]. Lymphocytic infiltration increases cellularity and fibrosis adds collagen, and both stiffen the gland.

Focal disease is equally relevant. The parotid is the most frequent site of benign salivary gland tumors, accounting for up to 70% of them, and the second most frequent site of malignant ones. Yet conventional imaging and palpation, which is inherently subjective, struggle to distinguish tumors from healthy tissue or benign from malignant lesions [2].

### *4.6 Radiation-Induced Fibrosis after Head and Neck cancer Treatment*

Head and neck cancer ranked as the sixth most common malignancy worldwide in 2022, with more than 940,000 new cases [99]. Treatment frequently causes late complications: more than 94% of patients treated with concomitant chemoradiotherapy develop at least one late complication within five years [100]. Radiation-induced fibrosis is observed in more than half of patients between one and eight years after radiotherapy [12].

Fibrosis is a dynamic pathological process involving damage to parenchymal cells, remodeling of the stroma, and tissue contraction. It is marked by excessive deposition of collagen and other extracellular matrix components, and by the presence of atypical fibroblasts [100]. Within the irradiated field, it can affect muscle, fascia, skin and subcutaneous tissue, and the salivary glands. Drawing on Section 3, the added collagen stiffens the extracellular matrix and is likely to increase its nonlinearity, while contraction shortens the affected tissues. Clinically, RIF contributes to dysphagia and pain, alongside late toxicities such as xerostomia, trismus and atrophy. Xerostomia, or dry mouth, reflects the salivary gland damage noted in Section 4.5. Despite this burden, fibrosis remains difficult to quantify objectively. In clinical practice it is still assessed subjectively, from symptoms and physical examination [100].

Taken together, altered tissue mechanics is a common feature of a wide range of neck conditions, and it arises through two main routes. In chronic neck pain, cervicogenic headache and occupational loading, stiffness changes mainly through altered muscle activation, tone and neuromuscular control, with secondary changes in passive muscle such as fat infiltration and atrophy. In malignancy, salivary gland disease and radiation-induced fibrosis, stiffness changes through altered tissue composition, namely increased collagen, increased cellularity, and raised fluid or interstitial pressure. These conditions are common and carry substantial disability or morbidity, yet they are still assessed largely subjectively, through palpation, symptoms and physical examination. Objective measurement of neck tissue mechanics therefore has clear clinical relevance for diagnosis, disease characterization and treatment monitoring. Section 5 describes how neck biomechanics are currently assessed and where these methods fall short.

## 5. Current Approaches for Measuring Neck Biomechanics

The tissue properties described in Section 3, and their alteration by the conditions reviewed in Section 4, are not directly measured by current assessment methods. This section describes the two main approaches used today. Section 5.1 covers clinical examinations, which evaluates neck function through tests of mobility, muscle performance and sensorimotor control. Section 5.2 covers computational modeling, which estimates internal loads and tissue deformation that cannot be measured in vivo.

### *5.1 Clinical Examination of Neck Biomechanics*

Clinical assessments evaluate cervical mobility, muscle performance, and sensorimotor control. They describe how the neck moves, generates force, and maintains head position, helping identify functional impairments and monitor changes over time [101].

Cervical mobility is commonly assessed by measuring range of motion (ROM) during flexion, extension, lateral bending, and axial rotation. During active testing, the individual moves their head and neck; during passive testing, an examiner moves them while the individual remains relaxed. These movements can be measured in degrees using instruments such as the cervical range-of-motion (CROM) device, which has demonstrated validity and reliability [102]. More focused examinations assess mobility within specific cervical regions. Passive segmental mobility testing involves manually moving cervical segments to assess restriction and resistance. The cervical flexion-rotation test assesses upper cervical rotation, primarily at C1-C2, with the neck positioned in flexion [103]. Together, these examinations help identify where movement is limited. However, limited ROM alone does not establish greater structural stiffness. Quantifying stiffness requires measuring both the applied force or moment and the resulting displacement or rotation.

Muscle performance is assessed through strength, endurance, and motor-control testing. Isometric strength testing measures force as the individual pushes against a dynamometer while keeping the head stationary [104]. Endurance testing measures how long a prescribed position can be maintained. For example, the neck flexor endurance test records how long the individual can hold the head slightly lifted while maintaining a chin tuck [105]. The cranio-cervical flexion test provides a more focused assessment of deep cervical flexor function. With the individual lying on their back, a pressure sensor behind the neck guides gentle nodding movements through progressively higher pressure targets. The examiner assesses movement control and the ability to sustain each target, providing an indirect measure of deep flexor activation and endurance [106]. Together, these tests characterize active muscle function but do not directly measure passive tissue stiffness.

Sensorimotor assessment evaluates how accurately an individual senses and controls head position. One approach is cervical joint-position error testing, in which the individual moves the head away from a starting position and then attempts to return to it without visual feedback. The difference between the starting and final positions measures repositioning accuracy. Revel, et al. [107] found that individuals with cervical pain were less accurate at repositioning their heads than healthy participants. This test therefore provides information about head-position sense and control, complementing ROM and strength tests, which assess movement range and force-generating capacity.

### *5.2 Computational Assessment of Neck Biomechanics*

Computational models complement clinical examinations by estimating internal loads and tissue deformation that are difficult or infeasible to measure in vivo. Finite element (FE) models divide anatomical geometry into elements and assign material properties, tissue interactions, and boundary conditions [108-110]. Depending on the structures represented, they can estimate intervertebral motion, disc deformation, ligament forces, and tissue stress and strain.

FE models distinguish passive tissue resistance from active muscle force generation. Vertebrae, discs, cartilage, ligaments, and joint capsules transmit loads and resist deformation through their assigned material properties and contact interactions. Passive muscle components represent resistance to stretch without activation. Active muscle contributions can be incorporated by muscle activation levels or prescribed forces applied along muscle paths. Prescribed forces impose a chosen loading condition but do not predict muscle recruitment. Simulating changes in recruitment requires a formulation that specifies activation over time; reproducing reflex responses additionally requires a feedback controller.

Model predictions depend on anatomical geometry, material properties, contacts, and loading assumptions. Numerical verification, including mesh-convergence assessment, evaluates whether the solution is precise. Validation compares predictions with experimental measurements under corresponding conditions. Computational results should therefore be interpreted according to the loading conditions and outputs for which the model has been evaluated.

Both approaches therefore leave a gap. Clinical tests describe how the neck functions but not the mechanical properties of individual tissues. Computational models depend on material properties that are largely derived from *ex vivo* cadaveric testing, which cannot capture muscle activation, *in vivo* hydration or differences between individuals. Elastography offers a way to measure tissue mechanical properties noninvasively in living subjects, and Sections 6 and 7 review its principles and applications in the neck.

## 6 Magnetic Resonance Elastography for the neck

Magnetic resonance elastography (MRE) is a noninvasive MRI-based imaging technique for tissue biomechanical properties, particularly stiffness [111]. Because it images the wave field in three dimensions and is not limited by acoustic access, MRE can assess tissues at any depth. It can also estimate mechanical properties beyond stiffness, such as viscosity and anisotropy. This section introduces the technique and reviews its application to the neck. Section 6.1 describes the three stages of an MRE examination: excitation (6.1.1), acquisition (6.1.2) and inversion (6.1.3). Section 6.2 introduces MRE biomarkers, from elasticity to viscosity, anisotropy, nonlinearity, porosity and hydrostatic pressure. Most of these have not yet been measured in the neck, so they are illustrated with studies of the liver, kidney, brain and skeletal muscle, and their potential relevance to the neck is noted. Finally, Section 6.3 reviews current applications of MRE in the neck.

### *6.1 Basic Principal of MRE*

An MRE examination has three stages: excitation, acquisition and inversion. In the excitation stage, an external vibration source applies harmonic vibration to the body surface, generating shear waves that propagate through the tissue. In the acquisition stage, the MRI scanner encodes the resulting tissue displacement into the phase of the MR signal, and the phase images are then converted into displacement maps. In the inversion stage, the mechanical properties are recovered from these maps. Each stage is described in turn below.

### *6.1.1 Excitation Methods*

Four main types of actuator are used to generate shear waves in MRE. The Lorentz-coil driver produces a cyclic force as its alternating current interacts with the scanner's static magnetic field [16, 112]. Such coils create their own magnetic field and thereby can distort the imaging signal, so clinical systems now mostly use metal-free pneumatic drivers [113]. In these, a loudspeaker outside the scanner room sends pressure waves through a tube to a passive pad placed on the body. Pneumatic drivers underpin clinical liver MRE, producing imageable shear waves throughout the upper abdomen even in obese patients and those with ascites. Piezoelectric drivers are used mainly for higher-frequency preclinical MRE [114]. A newer alternative, the gravitational transducer, rotates an eccentric mass to produce frequency-independent displacement with far lower harmonic distortion than pneumatic drivers (Runge et al., 2019). It has been demonstrated in the liver, kidney, prostate, breast and brain [115].

Drivers designed for the abdomen or brain are difficult to apply to the neck. Its targets are small, superficial and close to bone, and they lie beneath a curved surface whose shape varies between individuals. Neck studies have therefore adapted existing drivers. Yeung et al. fitted a commercial liver transducer with interchangeable headrest moulds [116], while other studies placed pads directly over the target muscle or gland [1, 22] (Figure 4; Section 6.3).

Excitation frequency is a fundamental excitation design parameter in MRE because it sets the shear wavelength. Higher frequencies produce shorter wavelengths, which resolve smaller structures, but they are attenuated more quickly and so penetrate less deeply into tissue. The frequency must therefore be matched to the size and depth of the target.

### *6.1.2 Acquisition Methods*

Acquisition strategies for neck MRE differ mainly in how motion encoding is generated and synchronized with the mechanical excitation, and in how many displacement components are captured. The most established approach uses a motion-sensitized spin-echo sequence with motion-encoding gradients (MEGs) phase-locked to the vibration, repetition and echo times set as integral multiples of the vibration period, and repeated acquisitions at equally spaced phase offsets across the mechanical cycle and along three orthogonal directions to recover the full displacement vector [3]. Echo-planar implementations of this spin-echo approach with trapezoidal MEGs support full 3D vector MRE, whereas the commercially available two-dimensional gradient-recalled echo (GRE) liver MRE sequence, which processes each slice independently using a single motion component, offers wider availability but generally overestimates stiffness relative to 3D vector reconstruction [2]. Where dedicated MEG capability is unavailable, the bipolar readout gradients of a conventional multi-echo gradient-echo sequence can be synchronized with the vibration so that successive echoes accumulate a progressively larger MEG-like phase shift, enabling MRE on conventional scanners at the cost of transverse relaxation and susceptibility-related phase wrapping in later echoes [117]. To make the acquisition faster, compressed sensing-based methods such as displacement-regularized compressed sensing method have been proposed by our group [118].

### *6.1.3 Inversion Methods*

The mechanical properties are not measured directly but inferred from their effect on the wave field, so elastography requires solving an inverse problem. Converting an imaged wave field into a stiffness value starts from the Cauchy–Navier equation of motion for a linear, isotropic elastic medium under time-harmonic excitation,

$$\nabla \cdot [\mu(\nabla u + (\nabla u)\top) + \lambda(\nabla \cdot u)I] = -\rho\omega^2 u \quad (2)$$

where u is the displacement vector, λ is the first Lamé parameter, ω = 2πf is the angular frequency of the excitation and I is the identity tensor. This equation reduces, under harmonic excitation and the near-incompressibility of soft tissue, to a Helmholtz equation for the shear wave alone [119]. The simplest direct inversion solves this voxel by voxel from a discrete Laplacian of the measured displacement field under an assumption of local homogeneity, which makes it sensitive to noise because differentiation amplifies measurement error, whereas combining data acquired at several driving frequencies increases the amount of information per unknown and improves stability[120]. Heterogeneous multifrequency direct inversion extends this by incorporating first-order stiffness gradients and combining a narrow band of frequencies, accommodating heterogeneous tissue within a fast, regularization-free solve[121]

Iterative model-based methods instead simulate wave propagation with finite elements and repeatedly update an assumed stiffness distribution until the computed and measured displacements agree; because they avoid differentiating the measured data they are less sensitive to noise, and they are typically solved over many small overlapping subzones [122-124]. To solve the challenge of slow convergence and local convergence, bi-convex based optimization have been proposed with structured sparsity using alternating direction method of multipliers [111, 125]. Most recently, deep learning approaches have been applied to wave inversion, including physics-informed neural networks that proved more robust to noise than conventional numerical methods[126, 127].

## *6.2 Candidate MRE Biomarkers*

Elasticity, quantified in MRE as the shear modulus μ (or storage modulus G′), describes tissue resistance to shear deformation and remains the primary biomarker reported by clinical elastography. It is the primary biomarker reported by clinical MRE, but MRE can also estimate viscosity, anisotropy, nonlinearity, porosity and hydrostatic pressure. Each of these reflects a different aspect of the tissue microstructure described in Section 3. The subsections below therefore illustrate each biomarker with studies of other organs, and end by noting its potential relevance to the neck.

### *6.2.1 Viscosity*

Viscosity captures the energy-dissipative component of tissue response to shear loading, represented in MRE as the loss modulus G″ within the complex shear modulus $G^* = G' + iG''$ (or, as a ratio, the loss angle φ, where tan φ = G″/G′), and reflects fluid content and viscous coupling between tissue compartments. Multifrequency MRE separates elastic and viscous behavior by exploiting the frequency dependence of measured stiffness: Kruse et al. showed in excised liver and kidney that shear stiffness increases systematically with frequency, consistent with a viscoelastic model from which shear modulus and shear viscosity can be derived independently [128]. In chronic viral hepatitis, three-dimensional multifrequency MRE (30–60 Hz) showed that the viscous parameter was effective for grading inflammatory activity [129].

### *6.2.2 Anisotropy*

Anisotropy describes the directional dependence of tissue stiffness caused by an aligned fibrous microstructure, such as muscle fascicles or white-matter tracts, and is typically modeled as transverse isotropy with distinct moduli parallel and perpendicular to the fiber direction. Early approaches supplied fiber orientation from diffusion tensor imaging (DTI) to the MRE inversion: in healthy lower-leg muscles, the shear modulus parallel to the fibers exceeded the perpendicular modulus by approximately 1.2–1.3-fold in the gastrocnemius, soleus, and tibialis anterior[130], and in fiber-reinforced phantoms and bovine muscle the MRE-derived anisotropy ratio agreed well with rheometry and distinguished different levels of fiber content[131].

### 6.2.3 Nonlinearity

Soft tissue does not remain equally stiff as it is deformed: at larger strains, or under an existing background stress, its stiffness itself changes. This nonlinearity can be quantified by applying a moderate static stress and tracking the resulting change in shear modulus, which recovers the Landau coefficient A. Two complementary strategies measure nonlinearity with elastography: detecting the higher harmonics that a finite-amplitude shear wave generates as it propagates through a nonlinear medium [132], and acoustoelasticity, which relates the change in shear wave speed under controlled uniaxial stress to the nonlinear shear modulus and has been validated with both ultrasound and MR elastography [133]. Nonlinearity is also a confound for conventional MRE, because physiological loading such as respiration or the cardiac cycle can bias apparent stiffness by up to twofold; incorporating knowledge of the tissue's rheology and loading state into the reconstruction removes this bias and recovers the intrinsic, unloaded stiffness[134].

### *6.2.4 Porosity*

Soft tissue often behaves less like a single solid and more like a sponge: a solid framework filled with fluid, where the two parts are mechanically coupled and share load as a wave passes through. This biphasic description was developed in geophysics for soils and has since been adapted for MRE, where simulations show it allows hydraulic permeability to be reconstructed alongside shear stiffness [135]. Poroelasticity sits alongside viscoelastic and anisotropic models as one of several extensions used to capture tissue mechanics beyond a single stiffness value. It introduces parameters absent from conventional viscoelastic MRE, namely the solid-matrix shear modulus, pore pressure, coupling density between solid and fluid, and porosity (fluid-volume fraction) [114]. Because a single-phase elastic model lumps the fluid's contribution into an "effective" shear modulus, it can systematically over- or underestimate the stiffness of the solid matrix wherever pore-pressure gradients develop [136]. A dedicated poroelastic inversion instead recovers the solid-matrix elastic properties together with an estimate of the pore-pressure distribution [137]. Porosity itself can be estimated noninvasively in the brain by exploiting the different T1 relaxation times of the solid and fluid compartments, an approach that also separates their shear wave fields, quantifies coupling density, and is proposed as a source of new imaging markers for research and diagnosis [138].

## *6.3 Current works on Neck MRE*

Compared with liver and brain MRE, MRE of the neck is an emerging field, but it has now been applied to several of the region's principal tissues. These are skeletal muscles such as the upper trapezius, the salivary and thyroid glands, cervical lymph nodes, and parotid tumors. The studies range from technical feasibility work, which establishes drivers, frequencies and normative values in healthy volunteers, to early clinical comparisons of lesions. Neck targets differ widely in size, depth and proximity to bone. As a result, each study has adapted its driver and tuned its excitation frequency to its target (Figure 4; Table 1). The studies also report different quantities (Section 6.2.1), so the quantity is stated with each value below.

Yeung et al. used 80 Hz for the parotid gland but 50 Hz for the thyroid to limit attenuation over the longer path from the driver and near bone, reporting storage moduli of 1.12 kPa for the normal parotid gland, 0.58 kPa for the normal thyroid, 0.66 kPa for a metastatic lymph node, and 0.17 kPa for a papillary thyroid carcinoma [3]. Ito et al., using a 3D-printed pneumatic pad on a conventional scanner, found 75 Hz to be the only frequency giving usable wave images in both the supraspinatus and trapezius at once, at which the trapezius was stiffer and more variable (7.26 ± 2.13 kPa) than the supraspinatus (4.16 ± 0.50 kPa), a difference the authors suggested might reflect the effects of mechanical neck pain [117]. In the upper trapezius, Hojo et al. found stiffness significantly higher when waves propagated along the fibers than across them at every frequency, with along-fiber stiffness increasing from 60 to 100 Hz, and recommended fiber-parallel excitation at 60 Hz, which gave a mean stiffness of approximately 4.2 kPa on both sides [139].

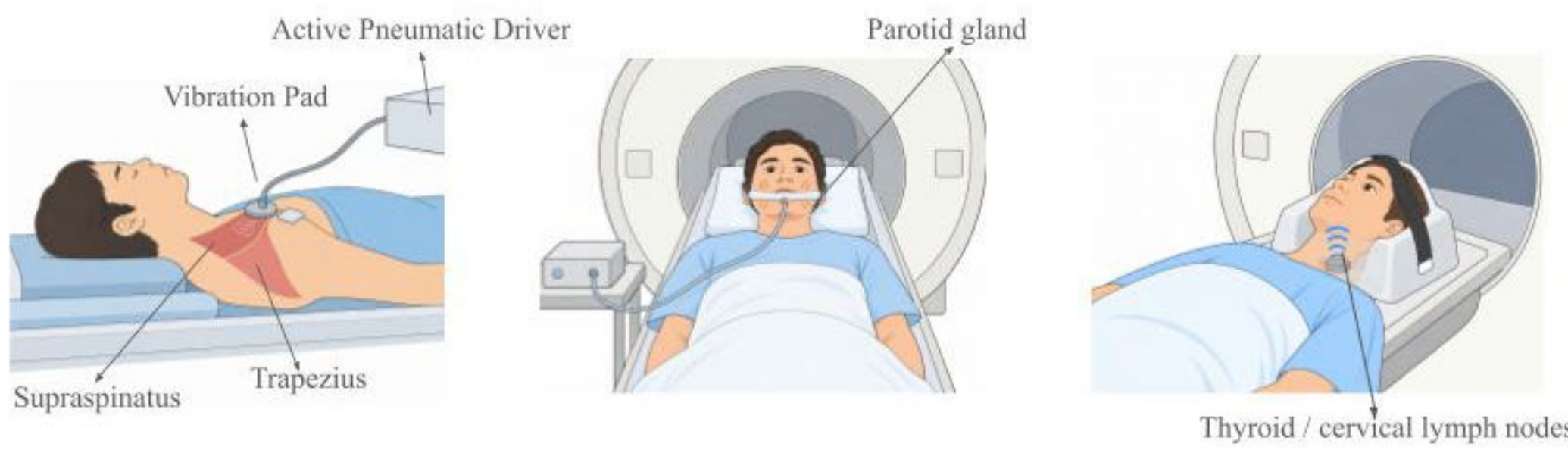

Figure 4 Examples of MRE driver designs used in the neck. (a) Pneumatic pad over the shoulder for the supraspinatus and trapezius, adapted from [1]. (b) Bilateral stethoscope-shaped driver for the parotid glands, adapted from [2]. (c) Driver in a headrest mould for the thyroid and cervical lymph nodes, adapted from [3].

Beyond muscle, DWI-based virtual MRE (Section 2.1.2) has been applied to cervical lymph nodes in head and neck squamous cell carcinoma, where metastatic nodes had a significantly higher mean elasticity than benign nodes (6.86 ± 0.82 vs 5.68 ± 0.94 kPa) [89]. Parotid MRE has not converged on a single driver or protocol. Elsholtz et al. introduced multifrequency waves (25–50 Hz) through the occiput with pressurized-air drivers and, without contacting the gland, reported a mean shear wave speed of 0.97 ± 0.13 m/s and loss angle of 0.59 ± 0.05 rad in healthy parotid glands [140], whereas Solak et al. positioned a commercial pneumatic liver driver directly over the gland at 60 Hz and reported a mean stiffness of approximately 1.2 kPa [141]. Because the gland is small, higher frequencies are needed for adequate resolution, which motivated Atamaniuk et al. to design a stethoscope-like driver with a pad over each gland; in a volunteer scanned with 3D MRE, this enabled imaging up to 120 Hz, over which parotid stiffness rose from 1.38–2.04 kPa(axial plane) at 60 Hz to 3.48–7.28 kPa(axial plane) at 120 Hz depending on whether brain- or liver-tuned inversion was used, and 2D MRE generally overestimated stiffness relative to 3D [2]. Applying this driver to patients, the same group found pleomorphic adenomas (3.72 ± 1.42 kPa) and mucoepidermoid carcinoma (3.41 kPa) stiffer than Warthin tumors (2.54 ± 0.70 kPa), and all tumors stiffer than healthy gland tissue (1.96 ± 0.44 kPa), although values overlapped between groups. [142]

Taken together, these studies show that MRE is feasible across the main tissues of the neck, including the upper trapezius, the parotid and thyroid glands, cervical lymph nodes, and parotid tumors. Progress has been driven largely by hardware and protocol adaptation: driver designs range from posterior-neck and occipital devices to pads placed directly over the muscle or gland, and excitation frequencies have been tuned to the size and depth of each structure. Early clinical studies have reported higher |G*| in parotid tumors than in healthy gland tissue, with values overlapping between tumor types, and higher shear modulus in metastatic than in benign lymph nodes using DWI-based virtual MRE, whereas the single papillary thyroid carcinoma examined had a lower G′ than normal thyroid tissue.

Table 1 Characteristics of MRE studies of the neck.

| Study | Target tissue | Driver / actuator | Excitation frequency | Study type / subjects | Key findings |
|---|---|---|---|---|---|
| [3] | Parotid gland, thyroid, cervical lymph nodes | Commercial liver-MRE transducer with interchangeable headrest mould and piston extension (posterior neck) | 80 Hz (parotid); 50 Hz (thyroid) | 6 healthy volunteers + 3 patients (3 metastatic nodes in 2 patients; 1 papillary thyroid carcinoma) | G′/G″: normal parotid 1.12/0.48 kPa; normal thyroid 0.58/0.42 kPa; metastatic node (mean of 3) 0.66/0.58 kPa; thyroid cancer 0.17/0.28 kPa |
| [1] | Supraspinatus and trapezius (simultaneous) | 3D-printed pneumatic vibration pad over the shoulder | 50–150 Hz (25 Hz steps); 75 Hz optimal | In vivo, human: 5 healthy volunteers | At 75 Hz shear stiffness: trapezius 7.26 ± 2.13 kPa; supraspinatus 4.16 ± 0.50 kPa |
| [139] | Upper trapezius | Pneumatic actuator placed in vertical (VAP, along-fiber) and horizontal (HAP, across-fiber) positions | 60, 80, 100 Hz | In vivo, human: 9 healthy volunteers | Shear stiffness = 4.2 kPa with VAP at 60 Hz (left 4.25 kPa, right 4.22 kPa); VAP > HAP at all frequencies; VAP stiffness increased with frequency, HAP did not |
| [140] | Parotid glands | Pressurized-air drivers against the occiput (no contact with gland) | 25, 30, 40, 50 Hz | In vivo, human: 20 healthy volunteers | Shear wave speed 0.97 ± 0.13 m/s; loss angle φ 0.59 ± 0.05 rad |
| [141] | Parotid gland | Commercial pneumatic liver driver positioned directly over the gland | 60 Hz | In vivo, human: 21 healthy volunteers | Mean shear stiffness ≈1.2 kPa (1.209 ± 0.240 kPa, reader 1; 1.146 ± 0.233 kPa, reader 2) |
| [2] | Parotid glands | Custom stethoscope-shaped passive driver with a pad over each gland | 3D: 60, 90, 120 Hz; 2D: 60, 80, 100, 120 Hz | In vivo, human: 3 healthy volunteers (1 with 3D MRE, 2 with 2D MRE) | 3D \|G*\| (both glands): 1.38 kPa (brain inversion) and 2.04 kPa (liver inversion) at 60 Hz; 3.48 and 7.28 kPa at 120 Hz (axial); 3.17 and 5.63 kPa at 120 Hz (coronal); 2D MRE generally overestimated stiffness vs 3D |
| [142] | Parotid tumors | Custom bilateral facial (stethoscope-type) driver | 80 Hz | In vivo, human: 14 patients with parotid tumors (healthy contralateral gland in 9) | Pleomorphic adenoma 3.72 ± 1.42 kPa; mucoepidermoid carcinoma 3.41 kPa; lymphoepithelial lesion 2.78 ± 0.33 kPa; Warthin tumor 2.54 ± 0.70 kPa; healthy gland 1.96 ± 0.44 kPa |
| [89] | Cervical lymph nodes (HNSCC) | DWI-based virtual MRE | N/A | In vivo, human: 16 metastatic nodes (10 patients) and 13 benign nodes (13 patients) | Mean shear modulas: metastatic 6.86 ± 0.82 kPa; benign 5.68 ± 0.94 kPa |

| Study | Target tissue | Driver / actuator | Excitation frequency | Study type / subjects | Key findings |
|---|---|---|---|---|---|
| [143] | Salivary gland tumors | DWI-based virtual MRE | N/A | In vivo, human: 34 tumors in 31 patients (17 benign, 6 Warthin, 11 malignant) | Median shear modulas: benign 1.9* kPa; malignant 3.94 kPa; Warthin 5.99 kPa |

*Note: G', storage modulus; G", loss modulus; |G*|, magnitude of the complex shear modulus; VAP/HAP, vertical/horizontal actuator position. *Values below 1.9 kPa, the lower limit of the calibration model, were recorded as 1.9 kPa.*

## 7 Ultrasound Elastography

### 7.1 Strain Elastography

Strain Elastography (SE) has been applied across a range of neck tissues to characterize relative mechanical properties and support the assessment of both normal and pathological tissue. Strain elastography estimates tissue stiffness by applying gentle external compression and tracking tissue displacement between pre and post-compression radio frequency echo signals [14]. Since softer tissue deforms more than stiffer tissue under equivalent stress, the resulting strain profile is mapped into a color or grayscale-coded elastogram depicting relative elasticity [14, 144-150].

#### 7.1.1 Neck Muscle Biomechanics

Strain-based assessment of neck muscle biomechanics can capture muscle deformation dynamically throughout movement. Using strain imaging on high-frame-rate ultrasound during a low-load neck extension exercise, Peolsson et al. found that nine women with chronic whiplash-associated disorders (WAD) showed greater shortening than matched controls in the trapezius and multifidus, but not in the three intermediate-depth extensors [151]. Using strain elastography with acoustic coupling gel as the reference, Yoshimura et al. found that self-stretching reduced upper trapezius strain-ratio in healthy adults in their 20s but not in those in their 40s [152].

#### 7.1.2 Thyroid Nodules

Strain elastography has shown variable diagnostic performance for thyroid nodules. In cytologically indeterminate (Thy-3) nodules, Cantisani et al. found that strain ratio outperformed the conventional ultrasound score (AUC 0.918 vs. 0.686). At a cutoff of 2.09, it achieved 90.6% sensitivity and 93% specificity for malignancy [153]. In contrast, a seven-centre study of 657 nodules from a broader population reported only 58% sensitivity and 78% specificity at a cutoff of 2.66 [154]. Surrounding thyroid tissue can be an unsuitable reference, for example in Hashimoto's thyroiditis. Krátky et al. therefore referenced nodule strain to the infrahyoid neck muscles instead. In 310 nodules, this approach achieved a negative predictive value of 92%, which rose to 100% when combined with conventional ultrasound risk factors [155]. In 310 thyroid nodules from 275 patients, the neck-muscle-referenced strain ratio achieved a negative predictive value of 92%, which increased to 100% when combined with conventional ultrasound risk factors [155]. Most recently, hard strain elastography was the only independent predictor of malignancy on histopathology. It was present in 79.3% of malignant nodules but only 7.7% of benign nodules [156].

### *7.1.3 Cervical Lymph Nodes and Other Soft Tissues*

Strain elastography of cervical lymph nodes follows the same diagnostic approach as in the thyroid. In 27 patients with enlarged nodes, B-mode assessment was fully sensitive (100%) but poorly specific (41.7%). A strain-ratio cutoff of ≥1.62 gave a better balance, with 93.3% sensitivity and 75% specificity [157]. In 123 abnormal nodes, an AREA% cutoff of 45.23% gave 84.3% sensitivity and 70.6% specificity [158]. Using the adjacent neck muscle as the reference tissue, Rawat et al. reported an AUC of 0.956 in 97 nodes, with 95.3% accuracy at a strain-ratio cutoff of 1.465 [159].

Strain elastography has also been applied beyond distinguishing benign from malignant nodes. In 120 patients with cervical tuberculous lymphadenitis, lesions with caseous necrosis (Type II) had the highest strain ratios. Elastography identified these lesions with 85.0% sensitivity and 76.3% specificity, and adding it to contrast-enhanced ultrasound raised sensitivity to 95.0%. Strain ratio also independently predicted poor treatment response at six months [160]. In 67 biopsy-proven cutaneous lesions, strain ratio completely separated malignant lesions (3.9–32.2) from benign ones (≤3.0) [161]. These results come from single cohorts and need validation, but they suggest that strain ratio can complement morphological assessment of superficial neck lesions.

## *7.2 Shear Wave Elastography*

Shear wave elastography (SWE) uses a focused acoustic radiation force, or "push beam", to generate shear waves within the tissue [15]. Ultrafast ultrasound imaging then tracks how these waves propagate, typically using cross-correlation, to measure the shear wave speed c [162-164]. Based on shear wave speed (Vs), tissue shear elastic modulus (μ) is calculated as $\mu = \rho Vs^2$ , where ρ is tissue density [162-164]. Compared with strain-based methods, SWE provides quantitative stiffness measurements and reduces dependence on operator-applied mechanical compression. These quantitative and relatively standardized measurements have enabled SWE to be increasingly applied across a range of neck tissues and pathologies.

### *7.2.1 Neck Muscle Biomechanics*

Shear wave elastography has characterized neck muscle stiffness across depth, activation and posture. In ten healthy volunteers, the resting shear modulus of the neck extensors increased from superficial to deep, from 7.7 kPa in the trapezius to 14.9 kPa in the multifidus. During isometric head lift, the semispinalis cervicis stiffened most, rather than the deepest muscle [165]. Posture also changes stiffness. Nuchal fascia shear modulus was higher in forward head posture than in neutral posture, on both the right (11.69 vs. 9.20 kPa) and left sides (11.28 vs. 8.93 kPa) [166]. Across eight muscles, posterior muscles were stiffest in flexion and softest in extension, whereas anterior muscles stiffened during deflection [20].

Evidence in chronic nonspecific neck pain is conflicting. Dieterich et al. found no difference in extensor stiffness between women with neck pain and controls, even though the pain group perceived their necks as stiffer [167]. Taş et al. found higher shear wave speed in patients in three of four muscles: the upper trapezius (3.2 vs. 2.8 m/s), levator scapulae (4.0 vs. 3.7 m/s) and sternocleidomastoid (2.5 vs. 2.2 m/s). The splenius capitis did not differ [9]. Neither study found that stiffness was related to pain or disability. In the largest study, of 174 patients and 142 controls, neither Young's modulus nor shear wave speed of the cervical multifidus distinguished patients from controls. Fatty infiltration measured on B-mode did (26.83% vs. 20.47%), which suggests that stiffness should complement morphological assessment rather than replace it [168].

Differences are clearer in more specific conditions. In cervicogenic headache, all four superficial extensors, including the splenius capitis, were stiffer on the headache side. The semispinalis cervicis separated patients from controls perfectly (AUC 1.000) [169]. In cervical myofascial pain syndrome, trapezius shear wave speed was higher than in patients without the syndrome (4.30 vs. 3.33 m/s). The AUC rose from 0.578 in mild pain to 0.983 in severe pain [170]. In neck somatic dysfunction, trapezius shear wave speed was higher (AUC 0.86) and fell after osteopathic manipulative treatment (1.96 to 1.58 m/s) [171]. After neck irradiation for head and neck cancer, muscle and fascial stiffness were about twice those of controls. For example, the median for the right sternocleidomastoid was 46.75 vs. 21.52 kPa, yet cervical range of motion was not reduced [172].

### *7.2.2 Thyroid Nodules*

Herman et al. reported a reference stiffness of 9.5 ± 3.6 kPa for the healthy thyroid in 128 volunteers [173]. In the largest diagnostic study, of 208 nodules, a cutoff of ≥26 kPa distinguished malignant from benign cytology (Bethesda V/VI vs. II) with an AUC of 0.876, 80% sensitivity and 81% specificity. Stiffness also remained an independent predictor of malignancy when compared with indeterminate cytology. It did not predict histopathological diagnosis, however; for that, hard strain elastography was the only independent predictor, suggesting that the two techniques provide complementary information [156]. In a pilot of six nodules, supersonic SWE classified all of them correctly. It also revealed punctate stiff foci corresponding to microcalcifications, some of which were not visible on B-mode imaging [174].

Thyroid SWE is, however, sensitive to probe pressure. In 19 healthy volunteers, stiffness rose non-linearly with probe force, from 16.7 kPa at 2 N to 29.9 kPa at 10 N. This is enough to move a benign nodule into the range typical of malignancy [21]. Across four precompression levels, stiffness increased in normal tissue, benign nodules and papillary carcinomas. The carcinomas stiffened most: by 75.4 kPa from baseline to high precompression, compared with 24.6 kPa for benign nodules and 10.8 kPa for normal tissue [175]. This strain hardening reflects the nonlinearity described in Section 6.2.4. Uncontrolled probe pressure can therefore contribute to variability in SWE values and diagnostic performance, and thyroid SWE requires standardized acquisition and appropriate reference values.

### *7.2.3 Cervical Lymph Nodes*

In healthy volunteers, Herman et al. reported a reference stiffness of 9.64 kPa for cervical lymph nodes [173]. Malignant nodes are consistently stiffer, but the reported cutoffs vary between studies. In 56 pathology-confirmed nodes, mean shear wave speed was 2.32 m/s in benign and 4.36 m/s in malignant nodes. A cutoff of 3.14 m/s gave 77.1% sensitivity and 85.7% specificity, and shear wave speed correlated strongly with the strain-based AREA% metric (r = 0.809) [176]. In a later head-to-head comparison in 123 nodes, SWE discriminated better than strain elastography overall (AUC 0.84 vs. 0.74; cutoff 3.05 m/s), although strain elastography was more sensitive [158]. In 97 nodes, by contrast, strain ratio slightly outperformed SWE (AUC 0.956 vs. 0.940). Even so, SWE reached 100% sensitivity at 2.195 m/s, which suggests it could help rule out malignancy [159]. SWE of the primary tumor can also predict nodal spread. In papillary thyroid carcinoma, an ARFI-derived area ratio above 1 predicted cervical lymph node metastasis with an AUC of 0.784. This rose to 0.803 when combined with conventional ultrasound features [177].

Surrounding tissue also affects nodal measurements. Values were higher when the probe was aligned parallel rather than perpendicular to the overlying sternocleidomastoid fibers. Stretching the neck increased stiffness only in the parallel orientation [17]. Together with the range of cutoffs reported (2.195–3.14 m/s), this shows that nodal SWE values depend on acquisition technique, and no single universal threshold has yet been established.

Table 2 Summary of strain elastography studies of the neck, grouped by tissue (neck muscle, thyroid, cervical lymph nodes and other soft tissues).

| Reference | Tissue Type and Dataset | Main experiment | Results and Findings |
|---|---|---|---|
| Peolsson, et al. [151] | 5 dorsal neck muscles (trapezius, splenius capitis, semispinalis capitis, semispinalis cervicis, and multifidus) in low-loaded neck extension exercise in 9 women with WAD and matched healthy controls. | To detect altered dorsal neck muscle behaviour in chronic WAD by quantifying frame-by-frame longitudinal strain. | WAD showed significantly greater shortening in the trapezius (extension phase) and in the multifidus (extension and return to neutral). The three intermediate-depth muscles showed no significant differences. The trapezius OPLS model had $Q^2Y = 0.18$ and the multifidus OPLS model had $Q^2Y = 0.27$. |
| [152] | Neck muscle- upper trapezius in 18 healthy adults in their 20s and 8 healthy adults in their 40s were included. | To measure upper trapezius muscle stiffness to assess the acute effects of self-stretching in healthy adults using strain-ratio approach<br>Also choroidal circulation was evaluated using laser speckle flowgraphy | In the 20-year-old group, trapezius muscle stiffness significantly decreased after self-stretching (baseline, 0.5 ± 0.1 vs. after stretching, 0.4 ± 0.1; $p < 0.001$).<br>In the 40-year-old group, no significant change was observed in muscle stiffness after stretching (baseline, 0.4 ± 0.1 vs. after stretching, 0.5 ± 0.2; $p = 0.179$) |
| [153] | Thyroid nodules - 315 patients with cytologically indeterminate (Thy-3) nodules; bi-centre university study | To determine whether strain ratio elastography improves the accuracy of conventional (colour-Doppler) ultrasound for characterizing indeterminate Thy-3 nodules and helps select patients for surgery. | Strain ratio significantly improved diagnostic performance over the conventional US score (AUC 0.918 vs. 0.686);<br>Higher strain ratio values were found in malignant nodules, with an optimum strain ratio cut-off of 2.09 at ROC analysis, and yielded 90.6% sensitivity and 93% specificity for malignancy. |
| [154] | Thyroid nodules - 602 patients with 657 nodules from seven German centres (prospective DEGUM multicentre study) | To evaluate the diagnostic performance of strain elastography for differentiating benign from malignant thyroid nodules in a multicentre setting. | Strain ratio optimal cutoff 2.66 was reported for malignancy: 58% sensitivity, 78% specificity, 71% accuracy |
| [155] | Thyroid nodules - 310 thyroid nodules in 275 patients | To test neck muscles and carotid artery content as alternative reference tissues for strain ratio, because surrounding thyroid tissue can be unsuitable (e.g., Hashimoto's thyroiditis, multinodular goitre, large nodules). | Strain ratio comparing the stiffness with neck muscles had a higher negative predictive value than elastographic score and conventional ultrasound (92 vs. 83 and 82 % respectively). The neck-muscle-referenced strain ratio achieved a negative predictive value of 100% when combined with conventional US. |
| [157] | Cervical lymph nodes in 27 patients (21–70 years) | To compare ultrasound elastography (elasticity score, strain ratio) with B-mode scoring for differentiating benign from malignant enlarged cervical lymph nodes. | B-mode: 100% sensitivity, 41.7% specificity, 74.1% accuracy. Strain ratio (cutoff ≥1.62): 93.3% sensitivity, 75% specificity, 85.2% accuracy. |
| [160] | Cervical lymph nodes: tuberculous lymphadenitis<br>120 patients with pathologically confirmed cervical tuberculous lymphadenitis (retrospective), pathological Types I–III | To evaluate contrast-enhanced ultrasound and ultrasound elastography for pathological classification and treatment-response prediction in cervical tuberculous lymphadenitis. | Strain elastography showed that Type II (caseous necrosis) nodes were the stiffest, with a mean strain ratio of 3.85, compared with 3.09 in Type III and 2.05 in Type I ($P < 0.001$). For identifying Type II, elastography had 85.0% sensitivity and 76.3% specificity. Adding elastography to CEUS raised sensitivity to 95.0%. |
| [161] | Skin (cutaneous lesions) - 55 patients with 67 biopsy-proven lesions (29 malignant, 38 benign) | To test whether strain ratio (lesion vs. adjacent normal skin) can discriminate benign from malignant cutaneous lesions, and to introduce the strain-ratio method. | Malignant lesions ranged from 3.9 to 32.2, and benign lesions ranged from 0.01 to 3.0. A cutoff between 3.0 and 3.9 would have given 100% sensitivity and specificity. Most cancers (79%) had ratios between 3.9 and 12.0. The lowest-ratio cancer (3.9) was a squamous cell carcinoma with central necrosis, which made its center softer. A basal cell carcinoma hidden beneath a benign seborrhoeic keratosis still produced a malignant-range ratio (17.65), even though it couldn't be seen on the skin surface. |

Table 3 Summary of shear wave elastography studies of the neck, grouped by tissue (neck muscle, thyroid, cervical lymph nodes)

| Reference | Tissue Type and Dataset | Main experiment | Results and Findings |
|---|---|---|---|
| [165] | Neck muscle - five layered extensors (trapezius, splenius capitis, semispinalis capitis, semispinalis cervicis, multifidus) in 10 healthy volunteers; prone, at rest and during graded isometric head-lift | To test whether the layered organization of the neck extensors (superficial to deep) is reflected in passive (rest) and active (head-lift) stiffness. | At rest, deeper muscles were stiffer, with multifidus highest (14.9 kPa). Next came semispinalis cervicis (9.5 kPa) and semispinalis capitis (8.9 kPa), then trapezius (7.7 kPa), with splenius capitis lowest (6.5 kPa). Semispinalis cervicis stiffened the most relative to rest, significantly more than multifidus. Trapezius and splenius capitis barely increased during the lift. |
| [166] | Neck muscle and fascia: nuchal fascia and upper cervical muscles (splenius capitis, semispinalis capitis, obliquus capitis superior and inferior) in 16 healthy adults; neutral vs. forward head posture | To determine the intra- and inter-rater reliability of SWE for nuchal fascia stiffness (and B-mode muscle thickness) and to compare neutral with forward head posture. | Nuchal fascia shear modulus was significantly higher in forward head posture than in neutral head posture on both the right (11.69 ± 1.99 vs. 9.20 ± 1.52 kPa, $P < 0.05$) and left sides (11.28 ± 2.26 vs. 8.93 ± 1.52 kPa, $P < 0.01$), Intra-rater reliability was good to excellent for nuchal fascia shear modulus (ICC 0.63-0.89) and all upper cervical muscle thicknesses (ICC 0.64-0.96), while inter-rater reliability was moderate to excellent for nuchal fascia shear modulus (ICC 0.54-0.82) and muscle thicknesses (ICC 0.48-0.88). |
| [20] | Neck muscle: 5 posterior (trapezius, splenius capitis, semispinalis capitis, semispinalis cervicis, multifidus) and 3 anterior (sternocleidomastoid, longus capitis, longus colli) in 100 healthy Chinese adults; multiple neck positions (including forward flexion, extension, deflection) | To assess thickness and stiffness of anterior and posterior neck muscles across neck positions and to evaluate measurement reliability, including the deep longus colli. | Posterior muscles were stiffest in forward flexion and softest in extension. Semispinalis capitis, measured 2.99, 4.61, and 2.35 m/s in neutral, flexion, and extension. All anterior muscles were stiffer in deflection; the sternocleidomastoid went from 2.67 to 4.13 m/s. Longus colli stiffness reliability was moderate-to-good (ICC 0.752-0.811). |
| [178] | Neck muscle: longus colli (deep anterior cervical flexor) in 19 adults with mechanical neck pain (≥6 months) | To determine intra- and inter-examiner reliability of SWE-derived longus colli stiffness under a standardized acquisition protocol. | Mean longus colli shear-wave speeds of 5.97-6.24 m/s and shear modulus values of 108.3-115.8 kPa, with no significant differences between sexes or sides. Good-to-excellent inter and intra-examiner reliability ($ICC_{3,2}$ >0.818; $ICC_{3,1}$ >0.891) under a standardized acquisition protocol |
| [17] | Phantom - A block of porcine (pig) sirloin muscle, In vivo - Reactive cervical lymph nodes at level III, behind the sternocleidomastoid (SCM) muscle, in Six healthy volunteers | To evaluate how the anisotropy and the static stretch stress of the cervical musculature influence the measured shear modulus in a tissue-mimicking phantom and in | Phantom- MeanSM varied significantly with depth (38-43 kPa at 0.5-1 cm). Strain significantly increased meanSM in both orientations. Parallel: 47.54 → 89.23 kPa, Perpendicular: 42.87 → 99.94 kPa |
| | | cervical lymph nodes in vivo | In vivo - Parallel values were significantly higher than perpendicular values, with or without stretch.Neutral: 17.49 vs. 10.99 kPa, Stretched: 37.11 vs. 13.24 kPa<br><br>Stretching significantly increased meanSM only in the parallel orientation. Parallel: 17.49 → 37.11 kPa Perpendicular: 10.99 → 13.24 kPa (P=0.100, not significant) |
| [167] | Neck muscle - five extensor layers (trapezius to multifidus) in women with chronic nonspecific neck pain (>6 months) and perceived neck stiffness (n = 20) vs. asymptomatic women (n = 18); relaxed state and multiple active tasks | To compare neck extensor stiffness between women with chronic nonspecific neck pain who report stiffness and asymptomatic controls. | No significant difference in pooled or task-specific stiffness (pain median 11.6 kPa vs. control 13.3 kPa, P = .175) despite the patients' perceived stiffness; stiffness was not associated with pain or disability. |
| [9] | Neck muscle: upper trapezius, levator scapulae, splenius capitis, sternocleidomastoid in 35 patients with chronic neck pain (≥3 months) vs. 35 age-matched asymptomatic controls | To compare neck muscle stiffness between patients with chronic neck pain and controls and to relate stiffness to pain intensity and disability. | Higher stiffness in the upper trapezius, levator scapulae and sternocleidomastoid; splenius capitis did not differ. Stiffness was not correlated with pain or disability scores. SWV, m/s: Upper trapezius- 3.2 vs. 2.8, Sternocleidomastoid: 2.5 vs. 2.2, Levator scapulae: 4.0 vs. 3.7 |
| [169] | Neck muscle: superficial cervical extensors (trapezius, splenius capitis, semispinalis capitis, semispinalis cervicis) in 19 patients with cervicogenic headache (headache side vs. non-headache side) vs. 20 healthy controls | To determine whether superficial cervical extensor stiffness is increased in cervicogenic headache and how well it discriminates patients from controls. | Stiffness was significantly increased on the headache side in all four muscles, including splenius capitis. Diagnostic AUC: trapezius 0.766, splenius capitis 0.759, semispinalis capitis 0.964, semispinalis cervicis 1.000 (strongest). |
| [168] | Neck muscle: cervical multifidus (C4-C5 level); B-mode metrics plus SWE in 174 patients with chronic idiopathic neck pain (≥12 weeks) vs. 142 asymptomatic controls | To determine the diagnostic accuracy of B-mode (area, perimeter, echo intensity, fatty infiltration) and SWE (Young's modulus, shear wave speed) metrics for discriminating chronic | Young's modulus and shear wave speed did not differ between groups (AUC 0.576 and 0.591, poor discrimination). Only fatty infiltration was higher in patients ($p < 0.001$; AUC 0.723). |
| | | neck pain from asymptomatic individuals. | |
| [170] | Neck muscle: upper trapezius in 109 patients with right posterior neck pain: 51 with cervical myofascial pain syndrome (MPS; mild, moderate, severe pain) vs. 58 without MPS | To evaluate upper trapezius elasticity in cervical MPS and its diagnostic value, including the relationship with pain severity. | Trapezius SWVmean was higher in MPS (4.30 vs. 3.33 m/s, $p < 0.001$). Mild-pain MPS patients did not differ from controls (3.51 vs 3.33 m/s), Moderate-pain (4.58 m/s), Severe-pain (5.40 m/s), Diagnostic AUC increased with pain severity, from 0.578 (mild) to 0.983 (severe). |
| [171] | Neck muscle: upper trapezius in 158 adults with and without neck somatic dysfunction (osteopathic TART criteria); before vs. after osteopathic manipulative treatment (OMT) | To determine whether SWE can detect neck somatic dysfunction and objectively capture the effect of OMT. | Higher longitudinal trapezius SWV in somatic dysfunction (moderate diagnostic accuracy, AUC 0.86); SWV decreased significantly after OMT (1.96 to 1.58 m/s, $p < 0.001$). |
| [172] | Neck muscle, fascia and subcutaneous tissue: sternocleidomastoid region in 16 head-and-neck cancer patients after bilateral cervical irradiation (median 26 months; mild fibrosis) vs. 16 matched non-irradiated controls | To evaluate SWE for tissue-specific quantification of radiation-induced cervical fibrosis and its relation to cervical range of motion. | Muscle showed the largest effect, about 2× higher than controls: Left: 46.98 vs. 28.10 kPa, Right: 46.75 vs. 21.52 kPa. Fascia showed similar increases, about 2.0×:Left: 46.41 vs. 25.47 kPa, Right: 44.28 vs. 22.69 kPa. Subcutaneous tissue showed a smaller but still significant increase, about 1.4-1.6×:Left: 34.16 vs. 23.89 kPa, Right: 34.31 vs. 20.92 kPa |
| [173] | Sternocleidomastoid muscle (also thyroid, lymph nodes, salivary glands, masseter) in 128 healthy volunteers (ages 21-91); bilateral measurements | To establish reference SWE stiffness values of normal neck soft tissues and to examine the influence of age, sex and BMI. | Mean stiffness clustered around 9-11 kPa across tissues: sternocleidomastoid 9.89 kPa, thyroid 9.5 ± 3.6 kPa, lymph node 9.64 kPa. No sex effect; only a small decrease with age (sternocleidomastoid significant on the left, not the right). |
| [21] | Thyroid (normal gland) in 19 healthy volunteers without known thyroid disease; probe preload force varied (2 to 10 N) | To quantify how probe preload force changes thyroid SWE stiffness estimates | Stiffness increased non-linearly with preload at 2 N to 19.3 kPa at 4 N, 22.3 kPa at 6 N, 25.9 kPa at 8 N, and 29.9 kPa at 10 N. |
| [175] | Thyroid: normal parenchyma, benign hyperplastic nodules, papillary carcinomas in | To assess how precompression influences SWE elasticity of thyroid tissue and nodules and | Stiffness increased with precompression in all three tissue types; papillary carcinomas showed the greatest strain hardening (75.4 |
| | 35 normal thyroid, 55 benign hyperplastic nodules and 17 papillary thyroid cancers (96 subjects); four operator-applied precompression levels | whether it may affect diagnostic performance | kPa increase from baseline to high precompression, vs. 24.6 kPa for benign nodules and 10.8 kPa for normal tissue). |

Table 4 Summary of studies that applied both strain and shear wave elastography within the same study population.

| Authors and Year | Tissue Type and Dataset | Main experiment | Results and Findings |
|---|---|---|---|
| [174] | Thyroid nodules - Case series: six nodules in four patients | To compare supersonic shear wave elastography (SSWE) with strain elastography in the differential evaluation of thyroid nodules. | SSWE gave 1 true-positive and 5 true-negative results, whereas strain elastography gave 5 false-positive and 1 false-negative results. SSWE showed punctate stiff foci corresponding to microcalcifications in 4 nodules, some not visible on B-mode. |
| [176] | Cervical lymph nodes in 56 enlarged nodes, pathology-confirmed (21 benign, 35 malignant); real-time elastography and VTIQ | To quantitatively analyse enlarged cervical lymph nodes with elastography (strain-based AREA% and VTIQ shear wave velocity) for benign vs. malignant differentiation | Mean SWV - 2.32 m/s (benign) vs. 4.36 m/s (malignant); a 3.14 m/s cutoff gave 77.1% sensitivity and 85.7% specificity. AREA% correlated strongly with SWV ($r = 0.809$). AREA% (strain elastography) was 61.31 ± 21.29% in malignant nodes and 45.03 ± 17.85% in benign nodes. |
| [177] | Cervical lymph node metastasis -Primary papillary thyroid carcinoma (used to predict cervical lymph node metastasis) in 203 patients with 222 histologically confirmed papillary thyroid carcinomas, with or without cervical lymph node metastasis (CLNM) | To investigate the value of combined conventional ultrasound (US), strain elastography (SE) and acoustic radiation force impulse (ARFI) elastography for prediction of cervical lymph node metastasis (CLNM) in papillary thyroid cancer (PTC) | SE was not significantly associated with metastasis; the ARFI-derived virtual touch tissue imaging area ratio (VAR) >1 was the strongest independent predictor (AUC 0.784), rising to 0.803 when combined with significant conventional US features. |
| [158] | Cervical lymph nodes in 123 abnormal nodes in 106 patients, pathology-confirmed (89 malignant, 34 benign) | To compare conventional ultrasound, strain elastography and shear wave elastography (Virtual Touch Tissue Imaging Quantification - VTIQ) for | Cutoff of 45.23% for SE (sensitivity 84.3%, specificity 70.6%) and 3.05 m/s for VTIQ (sensitivity 88.8%, specificity 73.5%). AUC for VTIQ was better than SE (0.84 vs. 0.74, $p = 0.005$). |
| | | detecting malignant cervical lymph nodes. | |
| [159] | Cervical lymph nodes - 97 nodes in patients aged 5-60 years | To assess the role of strain and SWE in differentiating benign from malignant cervical lymph nodes. | Malignant nodes had higher mean elasticity (72.06 vs. 9.23 kPa) and strain ratio (2.30 vs. 1.18). A strain ratio cutoff of 1.465 achieved 91.5% sensitivity and 99% specificity (AUC = 0.956). SWE velocity cutoff 2.195 m/s achieved 100% sensitivity and 85% specificity (AUC = 0.940). |
| [156] | Thyroid nodules - 208 nodules in 173 patients | To determine whether strain elastography (SE) and SWE provide complementary risk information beyond B-mode ultrasound for separating benign, indeterminate and malignant cytology. | Hard SE was the only independent predictor of malignant histopathological diagnosis in multivariable analysis, and was observed in 79.3% of malignant vs. 7.7% of benign nodules in the histopathological subgroup. SWE ≥26 kPa distinguished Bethesda V/VI from II (AUC 87.6%, 80% sensitivity, 81% specificity) and remained an independent predictor of malignancy when compared with indeterminate cytology. This did not extend to histopathology, where hard SE was the only independent predictor. |

## 8. Technical & Methodological Challenges

Neck elastography is limited less by any single technique than by a chain of dependencies, spanning wave generation, tissue physics, processing and clinical validation. These dependencies affect ultrasound elastography and MRE alike. This section considers each in turn and then proposes reporting recommendations.

**Wave generation and coupling.** Actuators and probes designed for large organs couple poorly to the small, curved and layered neck. In MRE, this has required custom moulds, pads and drivers for each target (Section 6.1.1). Skeletal muscle adds further difficulty, because fiber orientation, depth, thickness, activation and fibrosis all affect wave propagation. Characterizing anisotropy with additional encoding directions or diffusion imaging also lengthens acquisition [179]. In strain elastography, mobile neck structures can move out of the compression axis with arterial pulsation and breathing [90].

**Tissue physics: orientation, posture, muscle state and preload.** Because muscle is anisotropic, measured stiffness depends on imaging geometry in both modalities. Shear wave speed in five pericranial muscles was higher when scanning parallel rather than perpendicular to the fibers [180]. The same effect appeared in the trapezius with MRE [139] and in lymph nodes beneath the sternocleidomastoid with SWE [181]. Posture and muscle state are equally strong confounders.

- Posterior and anterior neck muscles change stiffness with head position [20, 166]
- Extensor shear modulus differs between rest and contraction [165]
- Combined flexion and contralateral bending raise upper trapezius and splenius capitis shear moduli above rest [182]

- In computer work, chair recline and head position alter trapezius stiffness, and the effect differs between men and women [8]

Probe pressure acts through tissue nonlinearity (Section 6.2.4). Within the clinical force range of 2–10 N, thyroid stiffness nearly doubled [21]. Papillary carcinomas stiffened more under precompression than benign or normal tissue [175]. Uncontrolled preload could therefore move a benign nodule into the malignant range.

**Processing and reported quantities.** In MRE, excitation frequency and processing choices account for much of the variability. Healthy parotid values obtained with low-frequency multifrequency excitation [183] and with a single-frequency liver driver [184] are not directly comparable. Within a single dataset, 2D acquisition gave higher values than 3D, and brain-tuned and liver-tuned inversions disagreed[22]. In SWE, converting wave speed to modulus assumes homogeneous, isotropic, purely elastic tissue, which muscle is not [15, 162]. The quantity reported also differs between modalities. Supersonic SWE reports Young's modulus, for example 9.0 ± 3.5 kPa for the healthy parotid and 9.5 ± 3.6 kPa for the thyroid [173]. MRE reports shear modulus, storage modulus or |G*|. Because $E \approx 3\mu$ in nearly incompressible tissue, kilopascal values cannot be compared across modalities without conversion. Strain elastography, meanwhile, is only semi-quantitative, and there is no consensus on the optimal strain-ratio cutoff [92].

**Table 5.** Recommended reporting items for neck elastography studies.

| Domain | Items to report |
|---|---|
| **System** | Modality and technique (strain elastography, SWE type, MRE sequence); vendor, system, probe or driver, and software version |
| **Reported quantity** | Exact quantity (E, μ, G′, G″, \|G*\|, φ, shear wave speed or strain ratio) and its units; for strain ratio, the reference tissue |
| **Excitation** | MRE: excitation frequency or frequencies, driver type and position. SWE: push and tracking settings where available |
| **Orientation** | Probe or wave-propagation direction relative to the muscle fibers |
| **Preload** | How probe pressure was controlled or measured (e.g., minimal-pressure technique, force sensor) |
| **Posture** | Body and head position, how it was standardized, and any support used |
| **Muscle state** | Rest or contraction level, and how it was verified (e.g., EMG, instruction, task) |
| **Acquisition** | 2D or 3D acquisition; number of motion-encoding directions (MRE); imaging depth; ROI size and placement |
| **Processing** | Inversion algorithm or reconstruction method; quality criteria and exclusions |
| **Reliability** | Number of repeated measurements; intra- and inter-examiner reliability in the study population |

Note: E, Young's modulus; μ, shear modulus; G′, storage modulus; G″, loss modulus; |G*|, magnitude of the complex shear modulus; φ, loss angle; EMG, electromyography; MRE, magnetic resonance elastography; ROI, region of interest; SWE, shear wave elastography

**Reliability and clinical validation.** SWE reliability is good to excellent for superficial muscles but lower for deep muscles (mean ICC 0.82 vs. 0.68, falling to 0.47 for the semispinalis capitis) [88]. Inter-examiner reliability is generally lower than intra-examiner reliability. Reliability evidence is also narrow. For example, good-to-excellent inter-examiner agreement for the sternocleidomastoid was shown only in asymptomatic young adults, at one supine site, on one platform and within one session [185]. No accepted normative ranges exist for cervical muscle stiffness, and clinical findings are inconsistent. In chronic nonspecific neck pain, some cohorts showed no stiffness differences [167], whereas another found higher stiffness in several muscles [9]. Cervical fascia stiffness was measured with excellent reliability (ICC $\geq$ 0.90) but did not differ between healthy participants and those with neck and low back pain [186]. Such discrepancies may reflect differences in participants, muscles, posture, contraction state and protocol.

**Reporting recommendations.** Neck elastography values are interpretable only when the factors above are specified. Table 5 lists the minimum information that studies should report to allow comparison across centers.

## 9. Conclusion

The neck is a coupled mechanical system in which passive tissues, active muscle, posture and loading act together, so its stiffness cannot be described by a single value. The mechanical properties of each tissue arise from its composition and structure, and they are viscoelastic, anisotropic, nonlinear and heterogeneous. Elastography therefore measures a regional tissue property, not whole-neck stiffness. That property depends on posture, muscle activation, excitation frequency and fiber orientation. It is nonetheless clinically relevant. Chronic neck pain, cervicogenic headache and occupational loading alter stiffness mainly through changes in muscle activation. Malignancy, salivary gland disease and radiation-induced fibrosis alter it through changes in tissue composition. These conditions are common and cause substantial morbidity, yet they are still assessed largely through palpation, symptoms and physical examination.

Ultrasound elastography is the more established modality in the neck. Strain and shear wave elastography have been applied widely to thyroid nodules, cervical lymph nodes and neck muscles, and the two techniques appear to provide complementary information. Malignant lesions are generally stiffer, though not universally. Muscle stiffness changes consistently in specific conditions such as cervicogenic headache and myofascial pain, but inconsistently in nonspecific neck pain. MRE of the neck is still emerging. It has been shown to be feasible in neck muscle, the salivary and thyroid glands, cervical lymph nodes and parotid tumors, mainly by adapting drivers and excitation frequencies to small, superficial targets. Early clinical results are encouraging but come from small cohorts. DWI-based virtual MRE should not be treated as equivalent to mechanical MRE. Beyond stiffness, MRE can also measure viscosity, anisotropy, nonlinearity and poroelastic properties. These remain almost unexplored in the neck, although each corresponds to a tissue property altered by the conditions reviewed here.

Across both modalities, reported values depend on how they are acquired and processed. The relevant factors include the modulus reported, excitation frequency, probe or actuator orientation, posture, muscle state, probe pressure, vendor and reconstruction method. Studies are therefore comparable only when these factors are standardized and reported. Future work should establish neck-specific protocols and normative values, and validate stiffness against clinical outcomes in larger cohorts. It should also compare ultrasound and MRE in the same subjects, and extend measurements beyond stiffness. In vivo measurements of neck tissue properties could also inform current cervical computational models, which rely largely on cadaveric and animal data.

References

[1] D. Ito *et al.*, "Simultaneous acquisition of magnetic resonance elastography of the supraspinatus and the trapezius muscles," *Magn Reson Imaging,* vol. 57, pp. 95-102, Apr 2019, doi: 10.1016/j.mri.2018.11.011.

[2] V. Atamaniuk *et al.*, "High-frequency shear wave MR elastography of parotid glands: custom driver design and preliminary results," *Scientific reports,* vol. 14, no. 1, p. 24496, 2024. [Online]. Available: https://www.nature.com/articles/s41598-024-75806-2.pdf.

[3] D. K. W. Yeung *et al.*, "MR elastography of the head and neck: driver design and initial results," *Magnetic resonance imaging,* vol. 31, no. 4, pp. 624--629, 2013.

[4] J. H. Bland and D. R. Boushey, "Anatomy and physiology of the cervical spine," in *Seminars in arthritis and rheumatism*, 1990, vol. 20, no. 1: Elsevier, pp. 1-20.

[5] S. Standring *et al.*, "Gray's anatomy: the anatomical basis of clinical practice," *American journal of neuroradiology,* vol. 26, no. 10, p. 2703, 2005.

[6] J. M. Barrett, C. McKinnon, and J. P. Callaghan, "Cervical spine joint loading with neck flexion," *Ergonomics,* vol. 63, no. 1, pp. 101-108, 2020, doi: 10.1080/00140139.2019.1677944.

[7] K. O. Akira Onda, "<Clinical features of neck and shoulder pain (Katakori) in Japanese hospital workers.pdf>," *Fukushima J. Med. Sci,* vol. 68, 2022.

[8] W. L. Wolff, C. M. Heinemann, J. M. Kartes, J. A. Ashton-Miller, and D. B. Lipps, "The influence of chair recline and head and neck position on upper trapezius activity and stiffness during seated computer work," *Applied Ergonomics,* vol. 117, p. 104227, 2024. [Online]. Available: https://www.sciencedirect.com/science/article/pii/S0003687024000048?via%3Dihub.

[9] S. Tas, F. Korkusuz, and Z. Erden, "Neck Muscle Stiffness in Participants With and Without Chronic Neck Pain: A Shear-Wave Elastography Study," *J Manipulative Physiol Ther,* vol. 41, no. 7, pp. 580-588, Sep 2018, doi: 10.1016/j.jmpt.2018.01.007.

[10] J. Ono, H. Nishi, S. Takashima, S. Soejima, R. Toya, and Y. Kumai, "Neck stiffness and its correlation with dysphagia after radiotherapy in head and neck cancer," *Auris Nasus Larynx,* vol. 52, no. 6, pp. 796-802, Dec 2025, doi: 10.1016/j.anl.2025.10.011.

[11] T. Rago *et al.*, "Low elasticity of thyroid nodules on ultrasound elastography is correlated with malignancy, degree of fibrosis, and high expression of galectin-3 and fibronectin-1," *Thyroid,* vol. 27, no. 1, pp. 103--110, 2017.

[12] B. Renslo *et al.*, "Quantification of Radiation-Induced Fibrosis in Head and Neck Cancer Patients Using Shear Wave Elastography," *Cureus,* vol. 16, no. 10, 2024.

[13] D. M. Elliott, H. R. Newman, M. N. Conner, C. L. Johnson, and E. J. Vresilovic, "Advances in mechanical assessments of in vivo human lumbar spine tissues with noninvasive imaging techniques," *NPJ Biomed Innov,* vol. 3, no. 1, p. 15, 2026/2// 2026, doi: 10.1038/s44385-026-00070-0.

[14] J. Ophir, I. Cespedes, H. Ponnekanti, Y. Yazdi, and X. Li, "Elastography: a quantitative method for imaging the elasticity of biological tissues," *Ultrason Imaging,* vol. 13, no. 2, pp. 111-34, Apr 1991, doi: 10.1177/016173469101300201.

[15] A. P. Sarvazyan, O. V. Rudenko, S. D. Swanson, J. B. Fowlkes, and S. Y. Emelianov, "Shear wave elasticity imaging: a new ultrasonic technology of medical diagnostics," *Ultrasound in medicine & biology,* vol. 24, no. 9, pp. 1419-1435, 1998.

[16] R. Muthupillai, D. J. Lomas, P. J. Rossman, J. F. Greenleaf, A. Manduca, and R. L. Ehman, "Magnetic resonance elastography by direct visualization of propagating acoustic strain waves," *science,* vol. 269, no. 5232, pp. 1854--1857, 1995.

[17] H. Y. Lee *et al.*, "Shear wave elastography using ultrasound: effects of anisotropy and stretch stress on a tissue phantom and in vivo reactive lymph nodes in the neck," *Ultrasonography,* vol. 36, no. 1, pp. 25-32, Jan 2017, doi: 10.14366/usg.16003.

[18] E. Hojo *et al.*, "Magnetic Resonance Elastography of Upper Trapezius Muscle," *NMR Biomed,* vol. 38, no. 4, p. e70007, Apr 2025, doi: 10.1002/nbm.70007.

[19] A. V. Dieterich *et al.*, "Shear wave elastography reveals different degrees of passive and active stiffness of the neck extensor muscles," *European Journal of Applied Physiology,* vol. 117, no. 1, pp. 171-178, 2017, doi: 10.1007/s00421-016-3509-5.

[20] R. Wang, X. Tang, and L. Qiu, "Assessment of neck muscle thickness and stiffness in Chinese healthy individuals by ultrasound," *Quant Imaging Med Surg,* vol. 15, no. 10, pp. 8910-8924, Oct 1 2025, doi: 10.21037/qims-2025-435.

[21] A. Ozturk *et al.*, "Variation of shear wave elastography with preload in the thyroid: quantitative validation," *Journal of Ultrasound in Medicine,* vol. 40, no. 4, pp. 779-786, 2021.

[22] V. Atamaniuk *et al.*, "High-frequency shear wave MR elastography of parotid glands: custom driver design and preliminary results," *Sci Rep,* vol. 14, no. 1, p. 24496, Oct 18 2024, doi: 10.1038/s41598-024-75806-2.

[23] J. T. Pruijssen, F. Schreuder, J. Wilbers, J. Kaanders, C. L. de Korte, and H. H. G. Hansen, "Performance evaluation of commercial and non-commercial shear wave elastography implementations for vascular applications," *Ultrasonics,* vol. 140, p. 107312, May 2024, doi: 10.1016/j.ultras.2024.107312.

[24] D. Klatt, S. Papazoglou, J. Braun, and I. Sack, "Viscoelasticity-based MR elastography of skeletal muscle," *Physics in Medicine and Biology,* vol. 55, no. 21, pp. 6445-6459, 2010, doi: 10.1088/0031-9155/55/21/007.

[25] H. Latorre-Ossa, J. L. Gennisson, E. De Brosses, and M. Tanter, "Quantitative imaging of nonlinear shear modulus by combining static elastography and shear wave elastography," *IEEE Trans Ultrason Ferroelectr Freq Control,* vol. 59, no. 4, pp. 833-9, Apr 2012, doi: 10.1109/TUFFC.2012.2262.

[26] M. A. Green, G. Geng, E. Qin, R. Sinkus, S. C. Gandevia, and L. E. Bilston, "Measuring anisotropic muscle stiffness properties using elastography," *NMR in Biomedicine,* vol. 26, no. 11, pp. 1387-1394, 2013, doi: 10.1002/nbm.2964.

[27] H. Bahreinizad, G. M. Paulon, L. Wei, and S. K. Chowdhury, "An MRI-derived head-neck finite element model," *Biomech Model Mechanobiol,* vol. 24, no. 6, pp. 2195-2215, Dec 2025, doi: 10.1007/s10237-025-02013-x.

[28] Ringleb, "Magnetic resonance elastography of healthy and pathologic skeletal muscle," *J Magn Reson Imaging,* 2007, doi: 10.1002/jmri.20817.

[29] Schmidt, "Magnetic resonance elastography of the lumbar spine: scoping review," *BMC Med Imaging,* 2025, doi: 10.1186/s12880-025-01662-9.

[30] M. Opara and Z. Kozinc, "Which muscles exhibit increased stiffness in people with chronic neck pain? A systematic review with meta-analysis," *Front Sports Act Living,* vol. 5, p. 1172514, 2023, doi: 10.3389/fspor.2023.1172514.

[31] Haueise, "Shear-wave elastography evidence map, musculoskeletal pain," *Clin Physiol Funct Imaging,* 2023, doi: 10.1111/cpf.12870.

[32] Singhal, "Finite element models of the cervical spinal cord," *NASS J,* 2023, doi: 10.1016/j.xnsj.2023.100246.

[33] Stanners, "Mechanical properties of the spinal cord (systematic review)," *Spine J,* 2024, doi: 10.1016/j.spinee.2024.02.022.

[34] N. Yoganandan, S. Kumaresan, and F. A. Pintar, "Biomechanics of the cervical spine Part 2. Cervical spine soft tissue responses and biomechanical modeling," *Clinical biomechanics,* vol. 16, no. 1, pp. 1-27, 2001.

[35] N. Bogduk and S. Mercer, "Biomechanics of the cervical spine. I: Normal kinematics," *Clinical Biomechanics,* vol. 15, no. 9, pp. 633-648, 2000, doi: 10.1016/s0268-0033(00)00034-6.

[36] H. Ansaripour, S. J. Ferguson, and M. Flohr, "In vitro biomechanics of the cervical spine: a systematic review," *Journal of Biomechanical Engineering,* vol. 144, no. 10, p. 100801, 2022.

[37] M. M. Panjabi, D. J. Summers, R. R. Pelker, T. Videman, G. E. Friedlaender, and W. O. Southwick, "Three-dimensional load-displacement curves due to froces on the cervical spine," *Journal of Orthopaedic Research,* vol. 4, no. 2, pp. 152-161, 1986, doi: 10.1002/jor.1100040203.

[38] R. Izzo, G. Guarnieri, G. Guglielmi, and M. Muto, "Biomechanics of the spine. Part I: Spinal stability," *European Journal of Radiology,* vol. 82, no. 1, pp. 118-126, 2013, doi: 10.1016/j.ejrad.2012.07.024.

[39] M. Panjabi *et al.*, "Three-Dimensional Movements of the Upper Cervical Spine," *Spine,* vol. 13, no. 7, pp. 726-730, 1988, doi: 10.1097/00007632-198807000-00003.

[40] C. Zhou *et al.*, "Intervertebral range of motion characteristics of normal cervical spinal segments (C0-T1) during in vivo neck motions," *Journal of Biomechanics,* vol. 98, p. 109418, 2020, doi: 10.1016/j.jbiomech.2019.109418.

[41] T. Ishii *et al.*, "Kinematics of the Subaxial Cervical Spine in Rotation In Vivo Three-Dimensional Analysis," *Spine,* vol. 29, no. 24, pp. 2826-2831, 2004, doi: 10.1097/01.brs.0000147806.31675.6b.

[42] T. Ishii *et al.*, "Kinematics of the Upper Cervical Spine in Rotation," *Spine,* vol. 29, no. 7, pp. E139-E144, 2004, doi: 10.1097/01.brs.0000116998.55056.3c.

[43] T. Ishii *et al.*, "Kinematics of the Cervical Spine in Lateral Bending," *Spine,* vol. 31, no. 2, pp. 155-160, 2006, doi: 10.1097/01.brs.0000195173.47334.1f.

[44] B. D. Stemper, N. Yoganandan, T. A. Gennarelli, and F. A. Pintar, "Localized cervical facet joint kinematics under physiological and whiplash loading," *Journal of Neurosurgery: Spine,* vol. 3, no. 6, pp. 471-476, 2005, doi: 10.3171/spi.2005.3.6.0471.

[45] T. Whyte, J. Barker, D. Cronin, G. A. Dumas, L.-P. Nolte, and P. Cripton, "Load-sharing and kinematics of the human cervical spine under multi-axial transverse shear loading: combined experimental and computational investigation," *Journal of biomechanical engineering,* vol. 143, no. 6, p. 061013, 2021.

[46] A. Trajkovski, S. Omerovic, S. Krasna, and I. Prebil, "Loading rate effect on mechanical properties of cervical spine ligaments," *Acta of Bioengineering and Biomechanics,* vol. 16, no. 3, pp. 13-20, 2014, doi: 10.5277/abb140302.

[47] K. M. Bell, Y. Yan, R. E. Debski, G. A. Sowa, J. D. Kang, and S. Tashman, "Influence of varying compressive loading methods on physiologic motion patterns in the cervical spine," *Journal of Biomechanics,* vol. 49, no. 2, pp. 167-172, 2016, doi: 10.1016/j.jbiomech.2015.11.045.

[48] D. D. Nevins, L. Zheng, and A. N. Vasavada, "Inter-individual variation in vertebral kinematics affects predictions of neck musculoskeletal models," *Journal of Biomechanics,* vol. 47, no. 13, pp. 3288-3294, 2014, doi: 10.1016/j.jbiomech.2014.08.017.

[49] W. Wei, S. Liao, S. Shi, J. Fei, Y. Wang, and C. Chen, "Straightened cervical lordosis causes stress concentration: a finite element model study," *Australasian Physical & Engineering Sciences in Medicine,* vol. 36, no. 1, pp. 27-33, 2013, doi: 10.1007/s13246-013-0182-4.

[50] G. P. Pal and H. H. Sherk, "The Vertical Stability of the Cervical Spine," *Spine,* vol. 13, no. 5, pp. 447-449, 1988, doi: 10.1097/00007632-198805000-00001.

[51] D. M. Skrzypiec, P. Pollintine, A. Przybyla, P. Dolan, and M. A. Adams, "The internal mechanical properties of cervical intervertebral discs as revealed by stress profilometry," *European Spine Journal,* vol. 16, no. 10, pp. 1701-1709, 2007.

[52] D. T. Reilly and A. H. Burstein, "The elastic and ultimate properties of compact bone tissue," *Journal of biomechanics,* vol. 8, no. 6, pp. 393-405, 1975.

[53] J. P. G. Urban and J. F. McMullin, "Swelling pressure of the inervertebral disc: Influence of proteoglycan and collagen contents," *Biorheology,* vol. 22, no. 2, pp. 145-157, 1985, doi: 10.3233/bir-1985-22205.

[54] V. C. Mow, S. Kuei, W. M. Lai, and C. G. Armstrong, "Biphasic creep and stress relaxation of articular cartilage in compression: theory and experiments," 1980.

[55] M. A. McLeod, R. E. Wilusz, and F. Guilak, "Depth-dependent anisotropy of the micromechanical properties of the extracellular and pericellular matrices of articular cartilage evaluated via atomic force microscopy," *Journal of biomechanics,* vol. 46, no. 3, pp. 586-592, 2013.

[56] J. Möller, L.-P. Nolte, H. Visarius, R. Willburger, J. Crisco, and M. Panjabi, "Viscoelasticity of the alar and transverse ligaments," *European Spine Journal,* vol. 1, no. 3, pp. 178-184, 1992.

[57] D. J. Lee and B. A. Winkelstein, "The failure response of the human cervical facet capsular ligament during facet joint retraction," *Journal of Biomechanics,* vol. 45, no. 14, pp. 2325-2329, 2012, doi: 10.1016/j.jbiomech.2012.07.015.

[58] H. D. Yahia, G. / Newman, N., "Structure-function relationship of human spinal ligaments," *Zeitschrift fur Mikroskopisch-anatomische Forschung,* vol. 104, no. 1, pp. 33-45, 1990.

[59] E. Ban, S. Zhang, V. Zarei, V. H. Barocas, B. A. Winkelstein, and C. R. Picu, "Collagen Organization in Facet Capsular Ligaments Varies With Spinal Region and With Ligament Deformation," *J Biomech Eng,* vol. 139, no. 7, pp. 0710091-9, Jul 1 2017, doi: 10.1115/1.4036019.

[60] S. P. Moroney, A. B. Schultz, J. A. Miller, and G. B. Andersson, "Load-displacement properties of lower cervical spine motion segments," *Journal of biomechanics,* vol. 21, no. 9, pp. 769-779, 1988.

[61] M. Simoneau, M. Denninger, and T. C. Hain, "Role of loading on head stability and effective neck stiffness and viscosity," *Journal of biomechanics,* vol. 41, no. 10, pp. 2097-2103, 2008.

[62] M. M. Panjabi, "The stabilizing system of the spine. Part I. Function, dysfunction, adaptation, and enhancement," *Journal of Spinal Disorders,* vol. 5, no. 4, pp. 383-389, 1992. [Online]. Available: https://pubmed.ncbi.nlm.nih.gov/1490034/.

[63] A. Race, N. D. Broom, and P. Robertson, "Effect of Loading Rate and Hydration on the Mechanical Properties of the Disc," *Spine,* vol. 25, no. 6, pp. 662-669, 2000, doi: 10.1097/00007632-200003150-00003.

[64] S. F. E. Mattucci, J. A. Moulton, N. Chandrashekar, and D. S. Cronin, "Strain rate dependent properties of younger human cervical spine ligaments," *Journal of the Mechanical Behavior of Biomedical Materials,* vol. 10, pp. 216-226, 2012, doi: 10.1016/j.jmbbm.2012.02.004.

[65] N. Yoganandan, F. Pintar, J. Butler, J. Reinartz, A. Sances, and S. J. Larson, "Dynamic Response of Human Cervical Spine Ligaments," *Spine,* vol. 14, no. 10, pp. 1102-1110, 1989, doi: 10.1097/00007632-198910000-00013.

[66] R. L. Lieber and B. I. Binder-Markey, "Biochemical and structural basis of the passive mechanical properties of whole skeletal muscle," *The Journal of physiology,* vol. 599, no. 16, pp. 3809-3823, 2021.

[67] G. A. Meyer and R. L. Lieber, "Elucidation of extracellular matrix mechanics from muscle fibers and fiber bundles," *Journal of biomechanics,* vol. 44, no. 4, pp. 771-773, 2011.

[68] P. P. Purslow, "Strain-induced reorientation of an intramuscular connective tissue network: implications for passive muscle elasticity," *Journal of biomechanics,* vol. 22, no. 1, pp. 21-31, 1989.

[69] D. E. Rassier and A. Månsson, "Mechanisms of myosin II force generation: insights from novel experimental techniques and approaches," *Physiological Reviews,* vol. 105, no. 1, pp. 1-93, 2025.

[70] W. L. Wolff, J. M. Leonardis, and D. B. Lipps, "The relationship between muscle activation and shear elastic modulus of the sternocleidomastoid muscle during 3-D torque production," *Journal of Electromyography and Kinesiology,* vol. 55, p. 102480, 2020, doi: 10.1016/j.jelekin.2020.102480.

[71] A. N. Vasavada, S. Li, and S. L. Delp, "Influence of Muscle Morphometry and Moment Arms on the Moment-Generating Capacity of Human Neck Muscles," *Spine,* vol. 23, no. 4, pp. 412-422, 1998, doi: 10.1097/00007632-199802150-00002.

[72] D. C. Ackland, J. S. Merritt, and M. G. Pandy, "Moment arms of the human neck muscles in flexion, bending and rotation," *Journal of Biomechanics,* vol. 44, no. 3, pp. 475-486, 2011, doi: 10.1016/j.jbiomech.2010.09.036.

[73] J. Schomacher, F. Petzke, and D. Falla, "Localised resistance selectively activates the semispinalis cervicis muscle in patients with neck pain," *Manual Therapy,* vol. 17, no. 6, pp. 544-548, 2012, doi: 10.1016/j.math.2012.05.012.

[74] Z. Wang, S. Petersson, R. Moreno, and R. Wang, "Anisotropic mechanical properties Quantification in skeletal muscle using magnetic resonance elastography and diffusion tensor imaging," *Journal of Biomechanics,* vol. 186, p. 112737, 2025, doi: 10.1016/j.jbiomech.2025.112737.

[75] G. D. O'Connell, S. Sen, and D. M. Elliott, "Human annulus fibrosus material properties from biaxial testing and constitutive modeling are altered with degeneration," *Biomechanics and Modeling in Mechanobiology,* vol. 11, no. 3-4, pp. 493-503, 2012, doi: 10.1007/s10237-011-0328-9.

[76] N. Buchweitz *et al.*, "Characterizing the Baseline Regional Biphasic Mechanical Properties of Cervical Intervertebral Discs," *Annals of Biomedical Engineering,* vol. 53, no. 9, pp. 2333-2345, 2025.

[77] S. P. Cohen, "Epidemiology, diagnosis, and treatment of neck pain," 2015.

[78] K. Ezzati *et al.*, "Prevalence of cervical myofascial pain syndrome and its correlation with the severity of pain and disability in patients with chronic non-specific neck pain," *Archives of Bone and Joint Surgery,* vol. 9, no. 2, p. 230, 2021. [Online]. Available: https://pmc.ncbi.nlm.nih.gov/articles/pmid/34026942/.

[79] S. P. Cohen and W. M. Hooten, "Advances in the diagnosis and management of neck pain," *Bmj,* vol. 358, 2017.

[80] A. Sawaf and A. Sawaf Sr, "Persistent Cervicothoracic Muscle Mechanical Stiffness as a Potential Time-Dependent Mechanism Linking Postural Loading to Chronic Neck and Upper Thoracic Pain: A Narrative Review," *Cureus,* vol. 18, no. 7, 2026.

[81] D. Falla, "Unravelling the complexity of muscle impairment in chronic neck pain," *Manual therapy,* vol. 9, no. 3, pp. 125--133, 2004.

[82] S. Shibasaki *et al.*, "Close relationships between neck and upper-back stiffness and transverse cervical artery flow velocity," *European Journal of Applied Physiology,* vol. 124, no. 6, pp. 1925--1931, 2024. [Online]. Available: https://link.springer.com/content/pdf/10.1007/s00421-024-05416-3.pdf.

[83] H. Ar and M. Gktepeli, "Ultrasonographic assessment of cervical and craniofacial muscle thickness in individuals with and without cervicogenic headache," *PeerJ,* vol. 14, p. e21285, 2026.

[84] B. Goadsby, "<Bartsch & Goadsby (2003) for the trigeminocervical complex.pdf>," 2003.

[85] D. M. Biondi, "<Cervicogenic headache - Diagnostic evaluation and treatment strategies.pdf>," *Current Pain and Headache Reports,* vol. 5, no. 4, pp. 361-368, 2001.

[86] J. Village and C. Trask, "Ergonomic analysis of postural and muscular loads to diagnostic sonographers," *International Journal of Industrial Ergonomics,* vol. 37, no. 9-10, pp. 781-789, 2007, doi: 10.1016/j.ergon.2007.06.005.

[87] J. Gremark Simonsen, A. Axmon, C. Nordander, and I. Arvidsson, "Neck and upper extremity pain in sonographers - Associations with occupational factors," *Appl Ergon,* vol. 58, pp. 245-253, Jan 2017, doi: 10.1016/j.apergo.2016.06.019.

[88] Y. Xie, L. Thomas, V. Johnston, and B. K. Coombes, "Cervical and axioscapular muscle stiffness measured with shear wave elastography: a comparison between different levels of work-related neck disability," *Journal of Electromyography and Kinesiology,* vol. 69, p. 102754, 2023.

[89] H. N. Jung, I. Ryoo, S. Suh, Y. H. Lee, and E. Kim, "Evaluating the elasticity of metastatic cervical lymph nodes in head and neck squamous cell carcinoma patients using DWI-based virtual MR elastography," *Magnetic Resonance in Medical Sciences,* vol. 23, no. 1, pp. 49--55, 2024. [Online]. Available: https://pmc.ncbi.nlm.nih.gov/articles/pmid/36529497/.

[90] K. S. S. Bhatia, Y. Y. P. Lee, E. H. Y. Yuen, and A. T. Ahuja, "Ultrasound elastography in the head and neck. Part I. Basic principles and practical aspects," *Cancer imaging,* vol. 13, no. 2, p. 253, 2013. [Online]. Available: https://pmc.ncbi.nlm.nih.gov/articles/pmid/23876352/.

[91] N. K. Altinbas and C. Yagci, "Sonoelastographic evaluation for benign neck lymph nodes and parathyroid lesions," *Journal of Ultrasonography,* vol. 18, no. 75, pp. 284--289, 2018. [Online]. Available: https://pmc.ncbi.nlm.nih.gov/articles/pmid/30763011/.

[92] A. S. McQueen and K. S. Bhatia, "Head and neck ultrasound: technical advances, novel applications and the role of elastography," *Clinical radiology,* vol. 73, no. 1, pp. 81--93, 2018. [Online]. Available: https://www.sciencedirect.com/science/article/pii/S0009926017304373?via%3Dihub.

[93] M. Pehlivan *et al.*, "Diagnostic role of ultrasound elastography on lymph node metastases in patients with head and neck cancer," *Brazilian journal of otorhinolaryngology,* vol. 85, no. 3, pp. 297--302, 2019. [Online]. Available: https://pmc.ncbi.nlm.nih.gov/articles/pmid/29933905/.

[94] C. Burusapat, W. Jarungroongruangchai, and M. Charoenpitakchai, "Prognostic factors of cervical node status in head and neck squamous cell carcinoma," *World J Surg Oncol,* vol. 13, p. 51, Feb 15 2015, doi: 10.1186/s12957-015-0460-6.

[95] L. C. Donaduzzi, F. De-Conto, L. S. Kuze, G. Rovani, M. E. Flores, and A. Pasqualotti, "Occurrence of contralateral lymph neck node metastasis in patients with squamous cell carcinoma of the oral cavity," *Journal of Clinical and Experimental Dentistry,* vol. 6, no. 3, p. e209, 2014. [Online]. Available: https://pmc.ncbi.nlm.nih.gov/articles/pmid/25136418/.

[96] Y. J. Choi, J. H. Lee, and J. H. Baek, "Ultrasound elastography for evaluation of cervical lymph nodes," *Ultrasonography,* vol. 34, no. 3, pp. 157--164, 2015. [Online]. Available: https://www.e-ultrasonography.org/upload/usg-15007.pdf.

[97] G. Maciel, C. S. Crowson, E. L. Matteson, and D. Cornec, "Prevalence of primary Sj \ o gren's syndrome in a US population-based cohort," *Arthritis care \& research,* vol. 69, no. 10, pp. 1612--1616, 2017.

[98] O. Unal, B. A. Dolek, A. Kor, E. S. Alcn, and S. Erten, "Diagnostic Performance of Parotid Shear-Wave Elastography for Predicting Histopathological Positivity in Patients with Suspected Primary Sj \ o gren's Syndrome," *Diagnostics,* vol. 16, no. 7, p. 1095, 2026. [Online]. Available: https://pmc.ncbi.nlm.nih.gov/articles/pmid/41975806/.

[99] J. Huang *et al.*, "Incidence, risk factors, and epidemiological trends of head and neck cancer: a global analysis," *MedComm,* vol. 6, no. 11, p. e70467, 2025. [Online]. Available: https://pmc.ncbi.nlm.nih.gov/articles/pmid/41244005/.

[100] E. C. Moloney, M. Brunner, A. J. Alexander, and J. Clark, "Quantifying fibrosis in head and neck cancer treatment: An overview," *Head Neck,* vol. 37, no. 8, pp. 1225-31, Aug 2015, doi: 10.1002/hed.23722.

[101] P. R. Blanpied *et al.*, "Neck pain: revision 2017: clinical practice guidelines linked to the international classification of functioning, disability and health from the orthopaedic section of the American Physical Therapy Association," *Journal of Orthopaedic & Sports Physical Therapy,* vol. 47, no. 7, pp. A1-A83, 2017.

[102] I. Audette, J.-P. Dumas, J. N. Côté, and S. J. De Serres, "Validity and between-day reliability of the cervical range of motion (CROM) device," *journal of orthopaedic & sports physical therapy,* vol. 40, no. 5, pp. 318-323, 2010.

[103] T. M. Hall, K. W. Robinson, O. Fujinawa, K. Akasaka, and E. A. Pyne, "Intertester reliability and diagnostic validity of the cervical flexion-rotation test," *Journal of manipulative and physiological therapeutics,* vol. 31, no. 4, pp. 293-300, 2008.

[104] K. T. Vannebo, V. M. Iversen, M. S. Fimland, and P. J. Mork, "Test-retest reliability of a handheld dynamometer for measurement of isometric cervical muscle strength," *Journal of back and musculoskeletal rehabilitation,* vol. 31, no. 3, pp. 557-565, 2018.

[105] T. Juul, H. Langberg, F. Enoch, and K. Søgaard, "The intra-and inter-rater reliability of five clinical muscle performance tests in patients with and without neck pain," *BMC musculoskeletal disorders,* vol. 14, no. 1, p. 339, 2013.

[106] G. A. Jull, S. P. O'leary, and D. L. Falla, "Clinical assessment of the deep cervical flexor muscles: the craniocervical flexion test," *Journal of manipulative and physiological therapeutics,* vol. 31, no. 7, pp. 525-533, 2008.

[107] M. Revel, C. Andre-Deshays, and M. Minguet, "Cervicocephalic kinesthetic sensibility in patients with cervical pain," *Archives of physical medicine and rehabilitation,* vol. 72, no. 5, pp. 288-291, 1991.

[108] M. B. Panzer, J. B. Fice, and D. S. Cronin, "Cervical spine response in frontal crash," *Medical engineering & physics,* vol. 33, no. 9, pp. 1147-1159, 2011.

[109] J. B. Barker and D. S. Cronin, "Multilevel validation of a male neck finite element model with active musculature," *Journal of biomechanical engineering,* vol. 143, no. 1, p. 011004, 2021.

[110] F. Li *et al.*, "A high-fidelity human cervical muscle finite element model for motion and injury studies," *Transportation safety and environment,* vol. 3, no. 4, p. tdab016, 2021.

[111] S. Mohammed *et al.*, "Model-based quantitative elasticity reconstruction using ADMM," *IEEE Transactions on Medical Imaging,* vol. 41, no. 11, pp. 3039--3052, 2022.

[112] M. Honarvar *et al.*, "Shielded Electromagnetic Transducer for Liver Magnetic Resonance Elastography," in *International Society for Magnetic Resonance in Medicine 2020 Annual Meeting & Exhibition (ISMRM)*, 2020.

[113] K. M. Pepin, C. L. Welle, F. F. Guglielmo, J. R. Dillman, and S. K. Venkatesh, "Magnetic resonance elastography of the liver: everything you need to know to get started," *Abdom Radiol (NY),* vol. 47, no. 1, pp. 94-114, Jan 2022, doi: 10.1007/s00261-021-03324-0.

[114] I. Sack, "Magnetic resonance elastography from fundamental soft-tissue mechanics to diagnostic imaging," *Nature Reviews Physics,* vol. 5, no. 1, pp. 25-42, 2022, doi: 10.1038/s42254-022-00543-2.

[115] O. I. Darwish *et al.*, "MR Elastography Using the Gravitational Transducer," *Sensors (Basel),* vol. 24, no. 24, Dec 17 2024, doi: 10.3390/s24248038.

[116] S. Rajagopalan *et al.*, "Enhanced visualization of ex-vivo ocular tissues using spatial-phase-resolved optical coherence microscopy," *Optics Letters,* vol. 50, no. 22, pp. 6959-6959, 2025/11// 2025, doi: 10.1364/ol.569766.

[117] D. Ito *et al.*, "Simultaneous acquisition of magnetic resonance elastography of the supraspinatus and the trapezius muscles," *Magnetic Resonance Imaging,* vol. 57, pp. 95--102, 2019.

[118] S. Mohammed, P. Kozlowski, and S. Salcudean, "Phase-regularized and displacement-regularized compressed sensing for fast magnetic resonance elastography," *NMR in Biomedicine,* vol. 36, no. 7, p. e4899, 2023.

[119] S. Papazoglou, U. Hamhaber, J. Braun, and I. Sack, "Algebraic Helmholtz inversion in planar magnetic resonance elastography," *Physics in Medicine \& Biology,* vol. 53, no. 12, pp. 3147--3158, 2008.

[120] D. Fovargue, D. Nordsletten, and R. Sinkus, "Stiffness reconstruction methods for MR elastography," *NMR Biomed,* vol. 31, no. 10, p. e3935, Oct 2018, doi: 10.1002/nbm.3935.

[121] E. Barnhill, P. J. Davies, C. Ariyurek, A. Fehlner, J. Braun, and I. Sack, "Heterogeneous Multifrequency Direct Inversion (HMDI) for magnetic resonance elastography with application to a clinical brain exam," *Medical image analysis,* vol. 46, pp. 180--188, 2018.

[122] E. E. W. Van Houten, M. I. Miga, J. B. Weaver, F. E. Kennedy, and K. D. Paulsen, "Three-dimensional subzone-based reconstruction algorithm for MR elastography," *Magnetic Resonance in Medicine: An Official Journal of the International Society for Magnetic Resonance in Medicine,* vol. 45, no. 5, pp. 827--837, 2001.

[123] M. D. J. McGarry *et al.*, "Suitability of poroelastic and viscoelastic mechanical models for high and low frequency MR elastography," *Medical physics,* vol. 42, no. 2, pp. 947--957, 2015. [Online]. Available: https://pmc.ncbi.nlm.nih.gov/articles/pmid/25652507/.

[124] S. Mohammed *et al.*, "Model-Based Quantitative Elasticity Reconstruction Using ADMM," *IEEE Trans Med Imaging,* vol. 41, no. 11, pp. 3039-3052, Nov 2022, doi: 10.1109/TMI.2022.3178072.

[125] S. Mohammed, M. Honarvar, P. Kozlowski, and S. Salcudean, "2D elasticity reconstruction with bi-convex alternating direction method of multipliers," in *2019 IEEE 16th International Symposium on Biomedical Imaging (ISBI 2019)*, 2019: IEEE, pp. 1683-1687.

[126] M. Ragoza and K. Batmanghelich, "Physics-informed neural networks for tissue elasticity reconstruction in magnetic resonance elastography," 2023.

[127] H. Bustin *et al.*, "Multifrequency Neural Network-based Wave Inversion in MR Elastography with Uncertainty Quantification," *Medical Image Analysis,* p. 104317, 2026.

[128] S. A. Kruse *et al.*, "Tissue characterization using magnetic resonance elastography: preliminary results," *Physics in Medicine \& Biology,* vol. 45, no. 6, pp. 1579--1590, 2000.

[129] Y. Wang *et al.*, "Viscoelastic parameters derived from multifrequency MR elastography for depicting hepatic fibrosis and inflammation in chronic viral hepatitis," *Insights into Imaging,* vol. 15, no. 1, p. 91, 2024. [Online]. Available: https://link.springer.com/content/pdf/10.1186/s13244-024-01652-5.pdf.

[130] M. A. Green, G. Geng, E. Qin, R. Sinkus, S. C. Gandevia, and L. E. Bilston, "Measuring anisotropic muscle stiffness properties using elastography," *NMR in Biomedicine,* vol. 26, no. 11, pp. 1387--1394, 2013.

[131] E. C. Qin *et al.*, "Combining MR elastography and diffusion tensor imaging for the assessment of anisotropic mechanical properties: a phantom study," *Journal of Magnetic Resonance Imaging,* vol. 37, no. 1, pp. 217--226, 2013.

[132] I. Sack, C. K. McGowan, A. Samani, C. Luginbuhl, W. Oakden, and D. B. Plewes, "Observation of nonlinear shear wave propagation using magnetic resonance elastography," *Magnetic Resonance in Medicine: An Official Journal of the International Society for Magnetic Resonance in Medicine,* vol. 52, no. 4, pp. 842--850, 2004.

[133] Pag, "Comparison of ultrasound elastography, magnetic resonance elastography and finite element model to quantify nonlinear shear modulus," *Physics in Medicine \& Biology,* vol. 68, no. 20, p. 205003, 2023.

[134] A. Capilnasiu *et al.*, "Magnetic resonance elastography in nonlinear viscoelastic materials under load: A. Capilnasiu et al.," *Biomechanics and modeling in mechanobiology,* vol. 18, no. 1, pp. 111--135, 2019. [Online]. Available: https://link.springer.com/content/pdf/10.1007/s10237-018-1072-1.pdf.

[135] D. R. Sowinski, M. D. J. McGarry, E. E. W. Van Houten, S. Gordon-Wylie, J. B. Weaver, and K. D. Paulsen, "Poroelasticity as a Model of Soft Tissue Structure: Hydraulic Permeability Reconstruction for Magnetic Resonance Elastography in Silico," *Front Phys,* vol. 8, Jan 2021, doi: 10.3389/fphy.2020.617582.

[136] P. R. Perriez, F. E. Kennedy, E. E. W. Van Houten, J. B. Weaver, and K. D. Paulsen, "Modeling of soft poroelastic tissue in time-harmonic MR elastography," *IEEE transactions on biomedical engineering,* vol. 56, no. 3, pp. 598--608, 2008.

[137] P. R. Perriez, F. E. Kennedy, E. E. W. Van Houten, J. B. Weaver, and K. D. Paulsen, "Magnetic resonance poroelastography: an algorithm for estimating the mechanical properties of fluid-saturated soft tissues," *IEEE transactions on medical imaging,* vol. 29, no. 3, pp. 746--755, 2010.

[138] L. Lilaj, T. Fischer, J. Guo, J. Braun, I. Sack, and S. Hirsch, "Separation of fluid and solid shear wave fields and quantification of coupling density by magnetic resonance poroelastography," *Magnetic Resonance in Medicine,* vol. 85, no. 3, pp. 1655--1668, 2021.

[139] E. Hojo *et al.*, "Magnetic Resonance Elastography of Upper Trapezius Muscle," *NMR in Biomedicine,* vol. 38, no. 4, p. e70007, 2025. [Online]. Available: https://pmc.ncbi.nlm.nih.gov/articles/pmid/40012134/.

[140] F. H. J. Elsholtz *et al.*, "Multifrequency magnetic resonance elastography-based tomoelastography of the parotid glands--feasibility and reference values," *Dentomaxillofacial Radiology,* vol. 51, no. 1, p. 20210337, 2022. [Online]. Available: https://pmc.ncbi.nlm.nih.gov/articles/pmid/34558305/.

[141] M. Solak, E. Kaba, M. Beyazal, M. eliker, and F. B. eliker, "Parotid gland magnetic resonance elastography feasibility study: clinical diagnostic potential and future perspectives as a radiological palpation method," *Diagnostics,* vol. 15, no. 18, p. 2351, 2025. [Online]. Available: https://pmc.ncbi.nlm.nih.gov/articles/pmid/41008723/.

[142] V. Atamaniuk *et al.*, "Magnetic resonance elastography for noninvasive evaluation of parotid tumors: a feasibility study," *Eur J Radiol,* vol. 202, p. 112970, Sep 2026, doi: 10.1016/j.ejrad.2026.112970.

[143] H. N. Jung, I. Ryoo, S. Suh, B. Kim, S.-H. You, and E. Kim, "Differentiation of salivary gland tumours using diffusion-weighted image-based virtual MR elastography: a pilot study," *Dentomaxillofacial Radiology,* vol. 53, no. 4, pp. 248--256, 2024. [Online]. Available: https://pmc.ncbi.nlm.nih.gov/articles/pmid/38502962/.

[144] I. Cespedes, J. Ophir, H. Ponnekanti, and N. Maklad, "Elastography: elasticity imaging using ultrasound with application to muscle and breast in vivo," *Ultrasonic imaging,* vol. 15, no. 2, pp. 73-88, 1993.

[145] M. Ashikuzzaman, A. Heroux, A. Tang, G. Cloutier, and H. Rivaz, "Displacement Tracking Techniques in Ultrasound Elastography: From Cross Correlation to Deep Learning," *IEEE Trans Ultrason Ferroelectr Freq Control,* vol. 71, no. 7, pp. 842-871, Jul 2024, doi: 10.1109/TUFFC.2024.3410671.

[146] M. Ashikuzzaman, C. J. Gauthier, and H. Rivaz, "Global Ultrasound Elastography in Spatial and Temporal Domains," *IEEE Trans Ultrason Ferroelectr Freq Control,* vol. 66, no. 5, pp. 876-887, May 2019, doi: 10.1109/TUFFC.2019.2903311.

[147] M. Ashikuzzaman *et al.*, "AutoSOUL: Deep Convolutional Neural Network for Autonomous Selection of Ultrasound Speckle Tracking Parameters," *IEEE Trans Ultrason,* vol. 73, no. 2, pp. 124-137, Feb 2026, doi: 10.1109/TUSON.2026.3651730.

[148] M. Ashikuzzaman, A. Sadeghi-Naini, A. Samani, and H. Rivaz, "Combining First- and Second-Order Continuity Constraints in Ultrasound Elastography," *IEEE Trans Ultrason Ferroelectr Freq Control,* vol. 68, no. 7, pp. 2407-2418, Jul 2021, doi: 10.1109/TUFFC.2021.3065884.

[149] M. Ashikuzzaman, A. K. Z. Tehrani, and H. Rivaz, "Exploiting Mechanics-Based Priors for Lateral Displacement Estimation in Ultrasound Elastography," *IEEE Trans Med Imaging,* vol. 42, no. 11, pp. 3307-3322, Nov 2023, doi: 10.1109/TMI.2023.3282542.

[150] M. Ashikuzzaman *et al.*, "MixTURE: L1-Norm-Based Mixed Second-Order Continuity in Strain Tensor Ultrasound Elastography," *IEEE Trans Ultrason Ferroelectr Freq Control,* vol. 71, no. 11, pp. 1389-1405, Nov 2024, doi: 10.1109/TUFFC.2024.3449815.

[151] A. Peolsson, G. Peterson, J. Trygg, and D. Nilsson, "Multivariate analysis of ultrasound-recorded dorsal strain sequences: Investigation of dynamic neck extensions in women with chronic whiplash associated disorders," *Sci Rep,* vol. 6, p. 30415, Aug 3 2016, doi: 10.1038/srep30415.

[152] M. Yoshimura, T. Taniguchi, T. Yoshitomi, and Y. Hashimoto, "Effects of Trapezius Muscle Self-Stretching on Muscle Stiffness and Choroidal Circulatory Dynamics: An Evaluation Using Ultrasound Strain Elastography and Laser Speckle Flowgraphy," *Tomography,* vol. 11, no. 7, Jun 25 2025, doi: 10.3390/tomography11070073.

[153] V. Cantisani *et al.*, "Strain ratio ultrasound elastography increases the accuracy of colour-Doppler ultrasound in the evaluation of Thy-3 nodules. A bi-centre university experience," *Eur Radiol,* vol. 26, no. 5, pp. 1441-9, May 2016, doi: 10.1007/s00330-015-3956-0.

[154] M. Friedrich-Rust *et al.*, "Evaluation of Strain Elastography for Differentiation of Thyroid Nodules: Results of a Prospective DEGUM Multicenter Study," *Ultraschall Med,* vol. 37, no. 3, pp. 262-70, Jun 2016, doi: 10.1055/s-0042-104647. Evaluation der Strain Elastografie fur die Differenzierung von Schilddrusenknoten: Ergebnisse einer prospektiven DEGUM Multizenter-Studie.

[155] J. Kratky, H. Vitkova, J. Bartakova, J. Lukas, and J. Jiskra, "Neck Muscles and Content of Carotid Artery as Reference Tissue for Strain Ratio - a Novel Approach to Improve the Diagnostic Performance of Thyroid Elastography?," *Exp Clin Endocrinol Diabetes,* vol. 124, no. 3, pp. 192-7, Mar 2016, doi: 10.1055/s-0035-1569359.

[156] K. A. A. Silveira, I. S. Barreto, I. M. B. Botelho, E. J. Pavin, A. Yamanaka, and D. E. Zantut-Wittmann, "Strain and Shear Wave Elastography in Thyroid Nodules: Tissue Stiffness as a Complementary Risk Marker for Differentiating Benign, Indeterminate, and Malignant Cytology," *Head Neck,* Aug 11 2026, doi: 10.1002/hed.70427.

[157] D. I. Hasan, A. F. Ahmed, R. Haggag, and A. e. m. Mohamed, "Ultrasound elastography in pathological enlarged cervical lymph nodes compared to histopathology," *The Egyptian Journal of Radiology and Nuclear Medicine,* vol. 47, no. 4, pp. 1349-1359, 2016, doi: 10.1016/j.ejrnm.2016.07.008.

[158] J. P. Zhang, Y. F. Xia, X. Han, and L. Shi, "Comparison of conventional ultrasonography and elastography ultrasound in detecting malignant cervical lymph nodes," *Pak J Med Sci,* vol. 41, no. 6, pp. 1689-1694, Jun 2025, doi: 10.12669/pjms.41.6.9215.

[159] A. Rawat, S. Mishra, N. A. Dikshit, P. Yadav, S. Kumar, and H. Nirwal, "A Comparative Evaluation of Strain and Shear Wave Ultrasound Elastography for Characterizing Cervical Lymphadenopathy," *Cureus,* vol. 17, no. 12, p. e100541, Dec 2025, doi: 10.7759/cureus.100541.

[160] X. Fu, X. Zhu, and X. Liu, "Clinical Value of Contrast-Enhanced Ultrasound and Ultrasound Elastography in the Pathological Classification and Treatment Response Prediction of Cervical Tuberculous Lymphadenitis," *Infect Drug Resist,* vol. 19, p. 610269, 2026, doi: 10.2147/IDR.S610269.

[161] B. Dasgeb, M. A. Morris, D. Mehregan, and E. L. Siegel, "Quantified ultrasound elastography in the assessment of cutaneous carcinoma," *Br J Radiol,* vol. 88, no. 1054, p. 20150344, Oct 2015, doi: 10.1259/bjr.20150344.

[162] F. Hug, K. Tucker, J. L. Gennisson, M. Tanter, and A. Nordez, "Elastography for Muscle Biomechanics: Toward the Estimation of Individual Muscle Force," *Exerc Sport Sci Rev,* vol. 43, no. 3, pp. 125-33, Jul 2015, doi: 10.1249/JES.0000000000000049.

[163] M. J. Alam, M. Ashikuzzaman, and M. A. L. Bell, "ShearMoFit: A Dual-Plane Ultrasound Shear Wave Motion Cleaning Technique," presented at the 2025 IEEE International Ultrasonics Symposium (IUS), 2025.

[164] M. Ashikuzzaman and M. A. Lediju Bell, "L1-Norm-Regularized Particle Motion Estimation in Ultrasound Shear Wave Elastography," presented at the 2024 IEEE Ultrasonics, Ferroelectrics, and Frequency Control Joint Symposium (UFFC-JS), 2024.

[165] A. V. Dieterich *et al.*, "Shear wave elastography reveals different degrees of passive and active stiffness of the neck extensor muscles," *Eur J Appl Physiol,* vol. 117, no. 1, pp. 171-178, Jan 2017, doi: 10.1007/s00421-016-3509-5.

[166] L. R. Liao, H. H. Pan, Y. X. Diao, Y. Xie, M. X. Liao, and Y. H. Xie, "Reliability of shear-wave elastography in assessing the stiffness of the nuchal fascia and the thickness of upper cervical muscles," *Med Eng Phys,* vol. 120, p. 104039, Oct 2023, doi: 10.1016/j.medengphy.2023.104039.

[167] A. V. Dieterich, U. S. Yavuz, F. Petzke, A. Nordez, and D. Falla, "Neck Muscle Stiffness Measured With Shear Wave Elastography in Women With Chronic Nonspecific Neck Pain," *J Orthop Sports Phys Ther,* vol. 50, no. 4, pp. 179-188, Apr 2020, doi: 10.2519/jospt.2020.8821.

[168] G. Plaza-Manzano *et al.*, "Diagnostic Accuracy of Ultrasound Imaging and Shear Wave Elastography to Discriminate Patients with Chronic Neck Pain from Asymptomatic Individuals," *Healthcare (Basel),* vol. 12, no. 19, Oct 5 2024, doi: 10.3390/healthcare12191987.

[169] L. Z. Lin *et al.*, "Increased Stiffness of the Superficial Cervical Extensor Muscles in Patients With Cervicogenic Headache: A Study Using Shear Wave Elastography," *Front Neurol,* vol. 13, p. 874643, 2022, doi: 10.3389/fneur.2022.874643.

[170] C. J. Hao *et al.*, "Upper trapezius muscle elasticity in cervical myofascial pain syndrome measured using real-time ultrasound shear-wave elastography," *Quant Imaging Med Surg,* vol. 13, no. 8, pp. 5168-5181, Aug 1 2023, doi: 10.21037/qims-22-797.

[171] J. Gao, E. Lynch, M. Coleman, J. Moore, D. Park, and B. Wilde, "Ultrasound shear wave elastography to assess neck somatic dysfunction and OMT effects," *J Osteopath Med,* vol. 125, no. 12, pp. 589-599, Dec 1 2025, doi: 10.1515/jom-2024-0236.

[172] H. Dapper, M. Waltenberger, S. U. Pigorsch, S. E. Combs, K. Bauermeister, and W. Bauermeister, "Shear-wave ultrasound elastography for tissue-specific quantification of radiation-induced cervical fibrosis," *Phys Imaging Radiat Oncol,* vol. 39, p. 101005, May 2026, doi: 10.1016/j.phro.2026.101005.

[173] J. Herman, Z. Sedlackova, J. Vachutka, T. Furst, R. Salzman, and J. Vomacka, "Shear wave elastography parameters of normal soft tissues of the neck," *Biomed Pap Med Fac Univ Palacky Olomouc Czech Repub,* vol. 161, no. 3, pp. 320-325, Sep 2017, doi: 10.5507/bp.2017.024.

[174] R. Z. Slapa *et al.*, "Shear wave elastography may add a new dimension to ultrasound evaluation of thyroid nodules: case series with comparative evaluation," *J Thyroid Res,* vol. 2012, p. 657147, 2012, doi: 10.1155/2012/657147.

[175] A. C. Lam, S. W. Pang, A. T. Ahuja, and K. S. Bhatia, "The influence of precompression on elasticity of thyroid nodules estimated by ultrasound shear wave elastography," *Eur Radiol,* vol. 26, no. 8, pp. 2845-52, Aug 2016, doi: 10.1007/s00330-015-4108-2.

[176] J. P. Zhang, H. Y. Liu, C. P. Ning, J. Chong, and Y. M. Sun, "Quantitative Analysis of Enlarged Cervical Lymph Nodes with Ultrasound Elastography," *Asian Pac J Cancer Prev,* vol. 16, no. 16, pp. 7291-4, 2015, doi: 10.7314/apjcp.2015.16.16.7291.

[177] J. M. Xu *et al.*, "Prediction of cervical lymph node metastasis in patients with papillary thyroid cancer using combined conventional ultrasound, strain elastography, and acoustic radiation force impulse (ARFI) elastography," *Eur Radiol,* vol. 26, no. 8, pp. 2611-22, Aug 2016, doi: 10.1007/s00330-015-4088-2.

[178] J. Izquierdo-Garcia *et al.*, "Testing the Reliability of a Procedure Using Shear-Wave Elastography for Measuring Longus Colli Muscle Stiffness," *Sensors (Basel),* vol. 26, no. 1, Dec 22 2025, doi: 10.3390/s26010065.

[179] S. Bensamoun, "<Current challenges of magnetic resonance elastography to quantify in vivo muscle.pdf>," 7 July 2025.

[180] C. Ewertsen, J. Carlsen, M. A. Perveez, and H. Schytz, "Reference values for shear wave elastography of neck and shoulder muscles in healthy individuals," *Ultrasound international open,* vol. 4, no. 01, pp. E23--E29, 2018.

[181] H. Y. Lee *et al.*, "Shear wave elastography using ultrasound: effects of anisotropy and stretch stress on a tissue phantom and in vivo reactive lymph nodes in the neck," *Ultrasonography,* vol. 36, no. 1, pp. 25--32, 2017. [Online]. Available: https://www.e-ultrasonography.org/upload/usg-16003.pdf.

[182] K. Yanase *et al.*, "Effective muscle elongation positions for the neck extensor muscles: An ultrasonic shear wave elastography study," *J Electromyogr Kinesiol,* vol. 60, p. 102569, Oct 2021, doi: 10.1016/j.jelekin.2021.102569.

[183] F. H. J. Elsholtz *et al.*, "Multifrequency magnetic resonance elastography-based tomoelastography of the parotid glands-feasibility and reference values," *Dentomaxillofac Radiol,* vol. 51, no. 1, p. 20210337, Jan 1 2022, doi: 10.1259/dmfr.20210337.

[184] M. Solak, E. Kaba, M. Beyazal, M. Celiker, and F. B. Celiker, "Parotid Gland Magnetic Resonance Elastography Feasibility Study: Clinical Diagnostic Potential and Future Perspectives as a Radiological Palpation Method," *Diagnostics (Basel),* vol. 15, no. 18, Sep 16 2025, doi: 10.3390/diagnostics15182351.

[185] G. Monclus-Diez *et al.*, "Reliability of Sternocleidomastoid Muscle Stiffness Assessment Using Shear-Wave Elastography Under a Standardized Protocol with Novice and Experienced Examiners: An Intra- and Inter-Examiner Reliability Study," *Medicina (Kaunas),* vol. 62, no. 2, Jan 27 2026, doi: 10.3390/medicina62020267.

[186] S. Zandi *et al.*, "Reliability of shear wave elastography for cervical and lumbar fascia stiffness and thickness in healthy and chronic neck and low back pain subjects," *Journal of Bodywork and Movement Therapies,* vol. 42, pp. 558--566, 2025.